\documentclass[useAMS,usenatbib,usegraphicx]{mn2eadapt}

\usepackage[latin1]{inputenc}
\usepackage[english]{babel}

\usepackage{amsmath}
\usepackage{amsfonts}
\usepackage{amssymb}
\usepackage{times}

\usepackage{xspace}

\usepackage{color}
\usepackage{rotating}
\usepackage{afterpage}
\usepackage{adjustbox}
\usepackage{xfrac}

\usepackage[colorlinks=false,dvips]{hyperref}
\hypersetup{pdfborder=0 0 0}

\usepackage{tikz}
\usetikzlibrary{trees}
\tikzstyle{level 1}=[level distance=2.5cm, sibling distance=4.0cm]
\tikzstyle{level 2}=[level distance=2.5cm, sibling distance=1.25cm]
\tikzstyle{bag} = [text width=4em, text centered]
\tikzstyle{end} = [circle, minimum width=3pt,fill, inner sep=0pt]

\newcommand{\eg}{e.g.\@\xspace}
\newcommand{\ie}{i.e.\@\xspace}
\newcommand{\cf}{cf.\@\xspace}

\newcommand{\MCh}{\ensuremath{M_\textrm{Ch}}}
\newcommand{\Msun}{\ensuremath{\textrm{M}_{\odot}}}

\newcommand{\Lsun}{\ensuremath{\textrm{L}_{\odot}}}
\newcommand{\kms}{\textrm{km\hspace{0.25em}s$^{-1}$}}

\newcommand{\OI}{\mbox{O\hspace{0.25em}{\sc i}}}

\newcommand{\NaI}{\mbox{Na\hspace{0.25em}{\sc i}}}
\newcommand{\KI}{\mbox{K\hspace{0.25em}{\sc i}}}
\newcommand{\MgII}{\mbox{Mg\hspace{0.25em}{\sc ii}}}

\newcommand{\SII}{\mbox{S\hspace{0.25em}{\sc ii}}}

\newcommand{\SiII}{\mbox{Si\hspace{0.25em}{\sc ii}}}

\newcommand{\CaII}{\mbox{Ca\hspace{0.25em}{\sc ii}}}
\newcommand{\TiII}{\mbox{Ti\hspace{0.25em}{\sc ii}}}

\newcommand{\FeII}{\mbox{Fe\hspace{0.25em}{\sc ii}}}
\newcommand{\FeIII}{\mbox{Fe\hspace{0.25em}{\sc iii}}}

\newcommand{\Fefs}{\ensuremath{^{56}}\textrm{Fe}}

\newcommand{\Cofs}{\ensuremath{^{56}}\textrm{Co}}
\newcommand{\Nifs}{\ensuremath{^{56}}\textrm{Ni}}

\newcommand{\Dm}{\ensuremath{\Delta m_{15}(B)}}

\newcommand{\quadbynine}{\hspace{0.1111em}}
\newcommand{\myto}{\hspace{0.18em}--\hspace{0.18em}}

\definecolor{myredder}{rgb}{1.0,0.0,0.0}

\begin{document}

\title[Spectral modelling of SNe Ia with(out) blueshifted \NaI\,D]{Type Ia supernovae with and without blueshifted narrow \NaI\,D lines -- how different is their structure?}

\author[Hachinger et al.]{S. Hachinger$^{1,2,3}$, F. K. Röpke$^{1,4}$, P. A. Mazzali$^{5,6}$, A. Gal-Yam$^{7}$, K. Maguire$^{8,9}$, \protect\vspace{0.3cm}\\
{\upshape\LARGE{\hspace{-0.13em}}M. Sullivan$^{10}$, S. Taubenberger$^{8,6}$, C. Ashall$^{5}$, H. Campbell$^{11}$, N. Elias-Rosa$^{12}$, U. Feindt$^{13}$, \protect\vspace{0.3cm}}\\
{\upshape\LARGE{\hspace{-0.13em}}L. Greggio$^{12}$, C. Inserra$^{9}$, M. Miluzio$^{12,14}$, S. J. Smartt$^{9}$, D. Young$^{9}$\protect\vspace{0.3cm}}\\
$^{1}$Institut f\"ur Theoretische Physik und Astrophysik, Universit\"at W\"urzburg, Emil-Fischer-Str. 31, 97074 W\"urzburg, Germany\\ 
$^{2}$Institut f\"ur Mathematik, Universit\"at W\"urzburg, Emil-Fischer-Str. 30, 97074 W\"urzburg, Germany\\ 
$^{3}$Leibniz Supercomputing Centre (LRZ), Bavarian Academy of Sciences and Humanities, Boltzmannstr. 1, 85748 Garching b. M\"unchen, Germany\\
$^{4}$Heidelberger Institut für Theoretische Studien, Schloss-Wolfsbrunnenweg 35, 69118 Heidelberg, Germany\\
$^{5}$Astrophysics Research Institute, Liverpool John Moores University, IC2 Liverpool Science Park, 146 Brownlow Hill, Liverpool, L3 5RF, UK\\
$^{6}$Max-Planck-Institut f\"ur Astrophysik, Karl-Schwarzschild-Str.\ 1, 85741 Garching, Germany\\
$^{7}$Benoziyo Center for Astrophysics, Weizmann Institute of Science, 76100 Rehovot, Israel\\
$^{8}$European Organisation for Astronomical Research in the Southern Hemisphere (ESO), Karl-Schwarzschild-Str. 2, 85748 Garching b. München, Germany\\
$^{9}$School of Mathematics and Physics, Queen's University Belfast, Belfast BT7 1NN, UK\\
$^{10}$Physics \& Astronomy, University of Southampton, Southampton, Hampshire SO17 1BJ, UK\\
$^{11}$Institute of Astronomy, University of Cambridge, Madingley Road, Cambridge CB3 0HA, UK\\
$^{12}$INAF - Osservatorio Astronomico di Padova, vicolo dell'Osservatorio 5, 35122 Padova, Italy\\
$^{13}$Oskar Klein Centre, Department of Physics, Stockholm University, Albanova University Center, 10691 Stockholm, Sweden\\
$^{14}$Instituto de Astrofisica de Canarias, C/ Vía Láctea, s/n, 38205, La Laguna, Tenerife, Spain\\
}

\date{ArXiv version 1.3 of manuscript accepted for publication in MNRAS.}
\pubyear{2017}
\volume{}
\pagerange{}

\maketitle

\begin{abstract}
In studies on intermediate- and high-resolution spectra of Type Ia supernovae (SNe Ia), some objects exhibit narrow \NaI\,D absorptions often blueshifted with respect to the rest wavelength within the host galaxy. The absence of these in other SNe Ia may reflect that the explosions have different progenitors: blueshifted \NaI\,D features might be explained by the outflows of `single-degenerate' systems (binaries of a white dwarf with a non-degenerate companion). In this work, we search for systematic differences among SNe Ia for which the \NaI\,D characteristics have been clearly established in previous studies. We perform an analysis of the chemical abundances in the outer ejecta of 13 `spectroscopically normal' SNe Ia (five of which show blueshifted Na lines), modelling time series of photospheric spectra with a radiative-transfer code. We find only moderate differences between `blueshifted-Na', `redshifted-Na' and `no-Na' SNe Ia, so that we can neither conclusively confirm a `one-scenario' nor a `two-scenario' theory for normal SNe Ia. Yet, some of the trends we see should be further studied using larger observed samples: Models for blueshifted-Na SNe tend to show higher photospheric velocities than no-Na SNe, corresponding to a higher opacity of the envelope. Consistently, blueshifted-Na SNe show hints of a somewhat larger iron-group content in the outer layers with respect to the no-Na subsample (and also to the redshifted-Na subsample). This agrees with earlier work where it was found that the light curves of no-Na SNe -- often appearing in elliptical galaxies -- are narrower, \ie\ decline more rapidly.
\end{abstract}


\section{Introduction}

Type Ia supernovae (SNe Ia) are one of the most well investigated but still poorly understood phenomena in astrophysics. The observed `homogeneity' of the light curves of  `normal' SNe Ia \citep{Hamuy1991a,Branch1993a,Phillips1993a,Goldhaber1998a} enables cosmologists to use them as standardisable candles \citep[\cf][]{Phillips1993a,Phillips1999a,Goldhaber2001a,Takanashi2008a}. The resulting Hubble diagrams have led to the remarkable conclusion that the expansion of the Universe is accelerating \citep[\eg][]{Riess1998a,Schmidt1998a,Perlmutter1999a,Riess2007a,Kessler2009a,Sullivan2011a}. 

Empirically, there are techniques to `standardise' or `calibrate' SNe Ia for cosmological purposes (\ie to predict their luminosity) with relatively little scatter \citep[\eg][]{Phillips1999a,Guy2007a,Jha2007a,Takanashi2008a,Wood-Vasey2008a}. However, SN cosmology and other fields of astronomy, being brought to new levels of accuracy, would require precise knowledge of the SN~Ia explosion mechanism \citep[\cf \eg][]{Foley2010a}, also predicting or excluding possible moderate drifts in the mean SN~Ia properties with look-back time, such as changes in mean luminosity. 
Despite a great effort to understand SNe Ia, this knowledge is still lacking -- it has been the matter of a long and ongoing debate \citep[\eg][]{Hillebrandt2000a,Greggio2005a,Howell2011a,Maoz2014a} whether the bulk of SNe Ia emerge from single-degenerate systems (where one C/O white dwarf with an extended companion star explodes) or from double-degenerate systems (where two white dwarfs merge, collide or one is accreted onto the other). Furthermore, there are open questions on how exactly the thermonuclear flame ignites and propagates.

The models which have probably been most extensively tested in order to explain SNe Ia \citep[for a review, see, e.g.,][]{Howell2011a} are (i) the `single-degenerate-Chandrasekhar-mass' model (presumably a `delayed detonation' of a C/O white dwarf before reaching its mass limit, \ie the Chandrasekhar mass \MCh\,$\sim$\,1.4\Msun), (ii) the `double-detonation' sub-\MCh\ model (a detonation of a less massive C/O white dwarf triggered by the detonation of an accreted helium shell), and (iii) the `double-degenerate' model (merger of two C/O white dwarfs).
Observations of SNe Ia in the UV, optical and IR have yielded constraints on the ejecta configuration and helped to demonstrate that recent explosion models may well explain the observed objects \citep[\eg$\!$][]{Baron2006a,Kasen2009a,Sim2010a,Kromer2013a,Hachinger2013a,Mazzali2014a}. However, only a few explosion models could definitely be excluded as for explaining normal SNe Ia [\eg pure deflagrations within single-degenerate systems, which match peculiar, 2002cx-like SNe Ia; see \citet[][]{Branch2004b,Sahu2008a,Jordan2012a,Kromer2013a}]. Thus, a search for further observational characteristics for verifying or falsifying promising models has begun. A promising group of methods is focused on finding direct signatures of the progenitor systems (\eg$\!$ \citealt[]{Patat2007a,Gilfanov2010a,Shen2013a,Silverman2013a,Kerzendorf2014a} -- for a review, see \citealt{Maoz2014a}).

One `smoking gun' signature of the progenitor system could be narrow, Doppler-shifted \NaI\,D absorption components from moving circumstellar material (CSM); see \eg\ \citet{Patat2007a,Simon2009a,Blondin2009a,Sternberg2011a,Foley2012c,Maguire2013a,Phillips2013a,Ferretti2016a}. \NaI\,D absorption components blueshifted with respect to the host galaxy indicate an outflow towards the observer -- material which may have been ejected by the progenitor system. This may be most easily accommodated in the single-degenerate scenario (\citealt{Patat2007a,Patat2011a,Booth2016a}; see, however, \citealt{Phillips2013a,Raskin2013a,Shen2013a,Soker2014a}), with a recurrent-nova progenitor system. In observed samples of SNe Ia, one sees red- as well as blueshifted \NaI\,D absorption systems, and many of these \citep[\eg$\!$][]{Maeda2016a} surely originate in the interstellar medium (ISM). However, samples of SNe Ia usually show more objects with blueshifted \NaI\,D components than redshifted ones (\eg \citealt{Sternberg2011a,Foley2012c,Maguire2013a}) -- suggesting\footnote{In addition, some of the objects exhibiting blueshifted \NaI\,D absorption show a time variability of the absorption lines, which has been interpreted as evidence of the ionisation conditions being modulated by the nearby SN (\citealt{Patat2007a,Simon2009a}, but see \eg \citealt{Chugai2008a}).} that some of the lines seen indeed have to do with the SN progenitors.

In this work, we test this idea, searching for differences in the mean physical properties of `blueshifted-Na', `single-' (not shifted, \cf Section \ref{sec:snsample-subsamples}) and `redshifted-Na' SNe, and SNe with no hints of \NaI\,D (`no-Na' SNe). We constrain ourselves to spectroscopically normal objects, and consider five blueshifted-Na, four redshifted-Na / single-Na and four no-Na SNe already classified (as for their CSM/ISM \NaI\,D properties) by \citet{Sternberg2011a,Foley2012c} and \citet{Maguire2013a}. The outer ejecta of these objects (i.e. the debris of the explosion -- not the CSM/ISM) are analysed by modelling the spectral time series of the SNe with a well-tested radiative transfer code \citep{Mazzali1993b,Lucy1999a,Mazzali2000a,Stehle2005a}. Afterwards, we compare the mean ejecta properties we find among the different `Na subsamples', focusing on the abundance structure of the explosion. In the abundances, we expect to possibly find differences from one Na subsample to another, just as those found between the single-degenerate delayed-detonation and double-degenerate violent-merger explosion models of \citet{Ropke2012a}. E.g.\@\xspace, the mass in iron-group elements in the outer layers of these models (above 8000\,\kms\ in velocity space) differs by 0.3\Msun, while the total mass contained in these layers is practically equal ($\sim$0.9\Msun).

Below (Section \ref{sec:method-data}), we first discuss our data selection in the context of the idea of this work. We also give details on our methods including the modelling technique. Afterwards, we show example models in Section \ref{sec:models}. In Section \ref{sec:analysis-discussion}, the results for our entire sample are analysed. Finally, we draw conclusions (Section \ref{sec:conclusions}).

\section{Data selection and modelling}
\label{sec:method-data}

This section discusses the basis of our work. We first recap some important studies on narrow \NaI\,D lines in high-resolution SN\,Ia spectra and illustrate what information high-resolution spectra provide in contrast to their low-resolution counterparts (Section \ref{sec:hiresvslores}). Afterwards, we describe our data sample and its treatment (Sections \ref{sec:snsample} \& \ref{sec:datahandling}). Finally, we lay out our method to model time series of low-resolution spectra (Section \ref{sec:modellingmethod}).

\subsection{High-resolution vs. low resolution spectra -- narrow \NaI\,D vs.\ broad ejecta features}
\label{sec:hiresvslores}

\begin{figure*}   
   \centering
   \hspace*{-0.3cm}
   \includegraphics[angle=0,width=10.5cm]{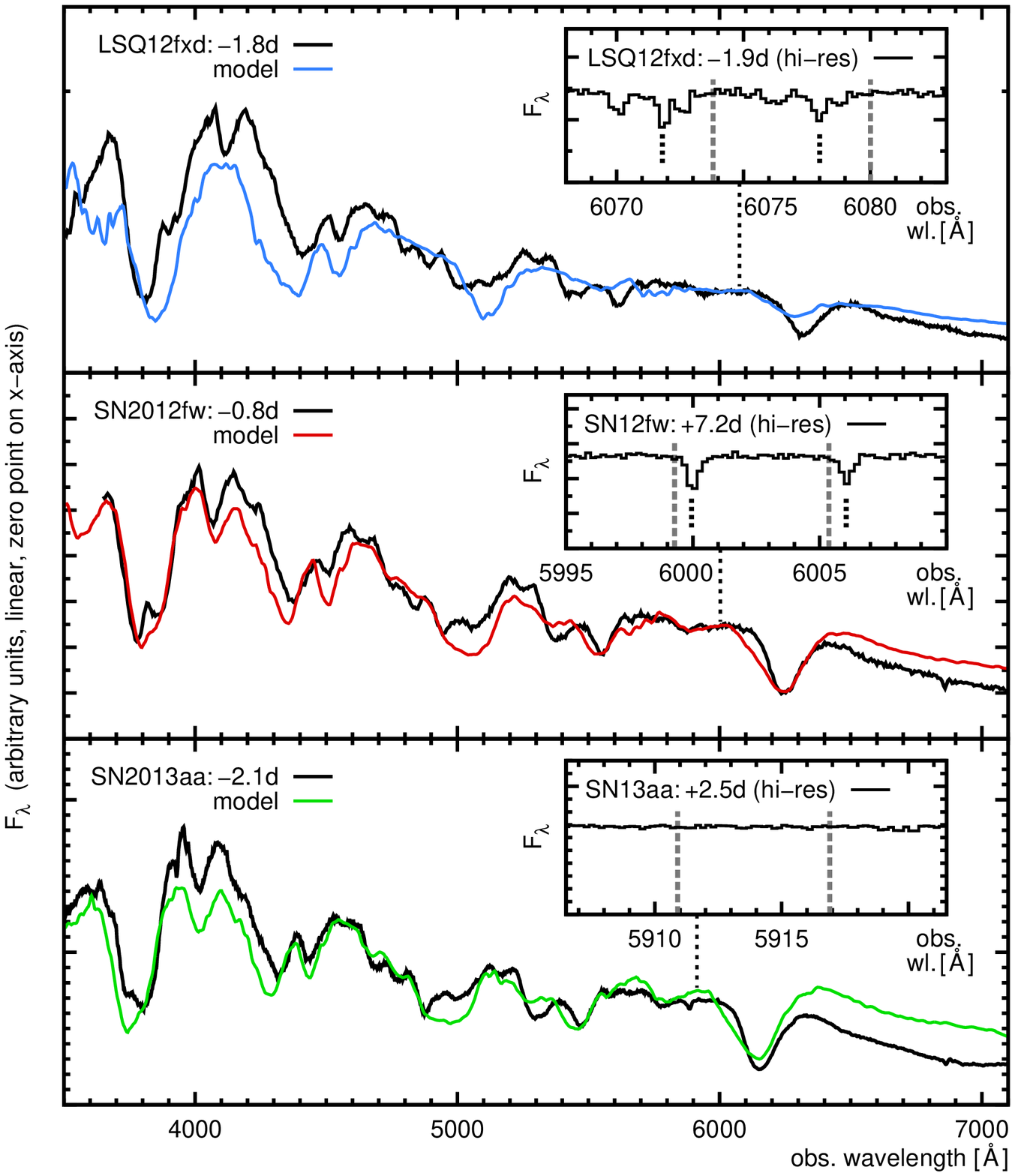}
   \caption{Typical spectra and models for SNe with blueshifted (LSQ\,12fxd), redshifted (SN\,2012fw) and absent (SN\,2013aa) narrow \NaI\,D lines. While \eg\ \citet{Sternberg2011a,Foley2012c,Maguire2013a} and \citet{Phillips2013a} have analysed \NaI\,D and other ISM/CSM features to establish the ISM/CSM properties, we model low-resolution SN spectra with their strongly Doppler-broadened lines produced by the ejecta (main panels, black lines; model spectra: grey/coloured lines). In the insets, we show 15\,\AA\ wide regions of high-resolution spectra for the SNe. The data shown were analysed by \citet{Maguire2013a} to examine the presence and possible wavelength shifts of narrow \NaI\,D features (\cf\ their Figures 1, 2, 3). We have marked the expected SN-site rest-frame wavelengths for \NaI\,D$_1$ and \NaI\,D$_2$ with grey dashes, the actual wavelengths of the minima with black dots.}
   \label{fig:loreshiresexamples}
\end{figure*}

Spectra of SNe Ia show strongly Doppler-broadened features produced by the fast-expanding ejecta (velocities in the order of 10,000\,\kms). These broad features can be analysed in low-resolution spectra. Their profiles (and the temporal evolution thereof) provide structural information about the SN material (composition, density, expansion velocities). On top of the `intrinsic' SN features, higher-resolution spectra show narrow features (velocities of a few hundred \kms) due to the presence of CSM or ISM. These features can provide information on the configuration of the progenitor system and/or the host galaxy.

In a pioneering work, \citet{Patat2007a} analysed the CSM/ISM \NaI\,D lines\footnote{The studies on \NaI\,D mentioned here also analyse different (often weaker) CSM/ISM features like \CaII\, H\,\&\,K or \KI\ $\lambda\lambda$7665,\,7699. In our study, we concentrate on the \NaI\,D information available for a relatively large number of objects.} of SN\,2006X in a time series of high-resolution spectra. They found that some CSM/ISM \NaI\,D components are blueshifted with respect to the rest frame of the SN, and that the strength of these components changes with time, which they interpreted as being caused by an ionisation effect from the SN [for a theoretical assessment of this, and possible caveats, see e.g. \citet{Chugai2008a} and \citet{Booth2016a}]. Estimating the properties of the CSM (extent, hydrogen mass, etc.), they concluded that the material may have originated from nova outbursts in a single-degenerate progenitor system. The variable-\NaI\,D phenomenon was later found to be present in more SN~Ia spectra \citep[\eg$\!$][]{Simon2009a,Blondin2009a,Stritzinger2010a}.

\NaI\,D properties in larger samples of SNe Ia have also been systematically analysed \citep{Sternberg2011a,Foley2012c,Maguire2013a,Phillips2013a}. These studies support the hypothesis that the lines sometimes arise from CSM. \citet{Sternberg2011a} reported an excess of SNe\,Ia showing blueshifted compared to redshifted \NaI\,D features with respect to the SN rest frame. This suggests that the material generating the features shows a preference of moving from the SN towards the observer. ISM clouds not related to the progenitor system would move randomly. Using larger samples of SNe Ia, \citet{Foley2012c} and \citet{Maguire2013a} confirmed this result. These studies also investigated correlations between \NaI\,D properties and SN properties (luminosity, expansion velocities, etc.) and found various possible hints the existence of two progenitor populations (one of them preferentially producing blueshifted-Na SNe\,Ia). \citet{Phillips2013a} found that SNe with blueshifted \NaI\,D tend to deviate from the normal correlation between ISM \NaI\,D column densities and the dust extinction (reddening). They show `too strong' \NaI\,D lines for `too little' extinction, suggesting that the material is CSM with a composition different from average ISM. 

\afterpage{\clearpage}
In this work, we perform a structural analysis of a SN Ia sample for which the properties of their \NaI\,D have previously been measured in moderate- and high-resolution spectra. To this purpose, we model the `intrinsic' SN features in time-series of spectra with a radiative-transfer code. This enables us to quantify differences in the structure of the SNe Ia linked to the presence or absence of blueshifted \NaI\,D absorption features. 
For our models reproducing the broad `intrinsic' SN features and their time evolution, it is sufficient to use low-resolution spectra (which have the advantage of well-sampled time series being available for many objects). Figure \ref{fig:loreshiresexamples} illustrates what information we extract from low-resolution spectra (main panels) as opposed to what the studies on \NaI\,D extract from high-resolution data (inserts). The Figure shows three example objects analysed in this work (LSQ\,12fxd, SN\,2012fw and SN\,2013aa), which show different \NaI\,D behaviours. Overplotted on the low-resolution SN spectra, we show our models discussed below (Sections \ref{sec:modellingmethod} and \ref{sec:models}). With the models, we aim at reproducing not only the data shown in the main panels of Figure \ref{fig:loreshiresexamples}, but the entire spectral time series of each SN around maximum light, inferring the ejecta characteristics (Section \ref{sec:modellingmethod}).

\begin{table*}
\begin{flushleft}
\caption{SNe grouped into the different subsamples. For each object we give $\Delta m_{15}(B)$, total extinction [for our moderate-redshift SNe: sum of the galactic and host-galaxy extinction values, $E(B-V)_\textrm{tot.}=E(B-V)_\textrm{MW}+E(B-V)_\textrm{host}$] and SN redshift ($z$) values assumed in the modelling process, and sources of the respective spectra.\label{tab:snsample}}
\begin{center}
\begin{tabular}{lllllll}
\hline
Object & $\!\!$ $\Delta m_{15}(B)^{a,c}$ $\!\!\!\!$ & $\!\!$ $E(B-V)_\textrm{tot.}^{b,c}$ $\!\!\!\!$ & host type$^{c}$ $\!\!\!\!$ & $z^{c}$ & references spectra & remarks\\
\hline
	\multicolumn{7}{c}{Blueshifted Na (no redshifted Na)}\\
\hline
	SN~2002bo & 1.2 & 0.38 & Sa  & 0.004 & \citet{Benetti2004a} & \parbox[t]{5.1cm}{Already modelled by \citet{Stehle2005a}; re-modelled according to the rules in Section \ref{sec:modellingmethod}.\vspace{0.09cm}} \\
	SN~2002ha & 1.2 & 0.23 & Sab & 0.014 & \citet{Blondin2012a} & \parbox[t]{5.1cm}{Calculated $E(B-V)_\textrm{host}$ of 0.14 from narrow \NaI\,D equivalent width$^d$ using the lower-limit relation of \citet{Turatto2003a}.\vspace{0.09cm}}  \\
	SN~2007le & 1.0 & 0.22$^e$ & Sc & 0.007 & \citet{Blondin2012a}  & \\
	SN~2008ec & 1.1 & 0.21 & Sa & 0.016 & \parbox[t]{2.6cm}{\citet{Silverman2012a}\,/$\!\!\!\!$ \citet{Milne2013a}} & \parbox[t]{5.1cm}{Assumed $E(B-V)_\textrm{host}$ value calculated by \citet{Milne2013a} from the peak photometry.\vspace{0.09cm}} \\
	LSQ\,12fxd$^f$ & 0.8$^h$ & 0.08 & Sc & 0.031 & \parbox[t]{2.38cm}{\citet{Maguire2013a}\,/$\!\!\!\!$ PESSTO} & \parbox[t]{5.1cm}{Inferred host extinction of 0.06 from \NaI\,D equivalent width$^d$ (see SN~2002ha; here: lower limit of various spectra.\vspace{0.08cm}}  \\
	
\hline
	\multicolumn{7}{c}{Redshifted Na (no blueshifted Na)}\\
\hline
	SN~2001el & 1.2 & 0.22 & Scd & 0.003 & \parbox[t]{2.55cm}{\citet{Wang2003a}; \citet{Mattila2005a}; \citet{Krisciunas2003a}$\!\!\!\!\!\!\!$ \vspace{0.09cm}} &   \\
	SN~2012fw$^g$ & 0.8$^h$ & 0.07 & S0/a & 0.019 & \parbox[t]{2.38cm}{\citet{Maguire2013a}\,/$\!\!\!\!$ PESSTO\vspace{0.09cm}} & \parbox[t]{5.1cm}{Inferred host extinction of 0.04 from \NaI\,D equivalent width$^d$ (see SN~2002ha).\vspace{0.09cm}}  \\
	SN~2012hr & 1.0$^h$ & 0.05 & Sbc & 0.007 & \parbox[t]{2.38cm}{\citet{Maguire2013a}\,/$\!\!\!\!$ PESSTO\vspace{0.09cm}} & \parbox[t]{5.1cm}{Inferred host extinction of 0.01 from \NaI\,D equivalent width$^d$ (see SN~2002ha).\vspace{0.09cm}} \\
\hline
	\multicolumn{7}{c}{Single (host-wavelength) Na}\\
\hline
	SN~2002cr & 1.2 & 0.08 & Scd & 0.009 & \citet{Blondin2012a} & \parbox[t]{5.1cm}{Inferred host extinction of 0.06 from \NaI\,D equivalent width$^d$ (see SN~2002ha).\vspace{0.09cm}} \\
\hline
	\multicolumn{7}{c}{No Na}\\
\hline
	SNF\,20080514-002 $\!\!\!\!\!\!\!$ & 1.4 & 0.03 & S0  & 0.022 & \parbox[t]{2.38cm}{\citet{Blondin2012a}\,/$\!\!\!\!$ \citet{Brown2012a}\vspace{0.09cm}} &  \\
	LSQ\,12dbr               & 0.9$^h$ & 0.06 & Irr & 0.020 &  \parbox[t]{2.38cm}{\citet{Maguire2013a}\,/$\!\!\!\!$ PESSTO} & \parbox[t]{5.1cm}{Assumed negligible host extinction because of absence of narrow \NaI\,D.\vspace{0.09cm}} \\
	SN~2012ht $\!\!\!\!\!\!\!$ & 1.4 & 0.03 & Irr  & 0.004 &  \parbox[t]{2.6cm}{\citet{Maguire2013a}\,/ PESSTO\,/$ \qquad \qquad $ \citet{Tomasella2014a}$^i\textrm{/}\!\!\!\!\!$ \citet{Yamanaka2014a}$\!\!\!\!$ \vspace{0.09cm}} & \parbox[t]{5.1cm}{Host extinction negligible \citep{Yamanaka2014a}.\vspace{0.09cm}} \\
	SN~2013aa $\!\!\!\!\!\!\!$ & 0.8$^h$ & 0.15 & Sc   & 0.004 &  \parbox[t]{2.38cm}{\citet{Maguire2013a}\,/$\!\!\!\!$ PESSTO\,/\,T.~Bohlsen$^j$} & \parbox[t]{5.1cm}{Assumed negligible host extinction because of absence of narrow \NaI\,D.\vspace{0.09cm}} \\
\hline
\end{tabular} 
\end{center}
\medskip
  $^{a}$ Reddening-free $B$-band decline within 15\,d directly after $B$ maximum \citep{Phillips1999a}.\\
  $^{b}$ Total dust extinction -- extinction within the host galaxy and the milky way can be added for low-redshift objects to obtain a total extinction $E(B-V)_\textrm{tot.}$.\\
  $^{c}$ Basic data [$\Delta m_{15}(B)$, $E(B-V)$, host type, $z$] for the SNe have been taken from \citet{Maguire2013a}, \citet{Foley2012c}, \citet{Sternberg2011a} and the cited papers for each SN, unless noted otherwise.\\
  $^{d}$ To the purpose of applying the relation of \citet{Turatto2003a}, which approximates the extinction based on narrow \NaI\,D features as seen in \textit{low-resolution} data, we measured the equivalent width of the narrow \NaI\,D feature at the host-galaxy rest wavelength in our low-resolution spectra whenever the line was measurable (including \NaI\,D components which may be due to the SN progenitor -- also this dust will redden the SN) and took the average. The uncertainty in the reddening introduced here has no major impact on our results, as all objects affected are relatively weakly reddened.\\
  $^{e}$ From \citet{Simon2009a} -- in their work, $E(B-V)_\textrm{tot.} = \textrm{0.27}$ with $R_V=\textrm{2.59}$. For consistence with the other SNe, we assume $R_V=\textrm{3.1}$ and a lower reddening of $E(B-V)_\textrm{tot.} = \textrm{0.22}$, for which our reddening law best reproduces the $A_B$ and $A_V$ values of \citet{Simon2009a} (with the same error each).\\
  $^{f}$ LSQ\,12fxd, with respect to the strongest \NaI\,D component, displays a blueshifted component as well as a redshifted component (slightly weaker and attached to the strongest component). It may thus have been classified as `symmetric' by \citet{Sternberg2011a}.\\
  $^{g}$ SN~2012fw displays only one strong \NaI\,D line extended somewhat more to the red. \citet{Sternberg2011a} may thus have classified it as `symmetric'.\\
  $^{h}$ Calculated from the stretch values of \citet{Maguire2013a}, using equation (5) of \citet{Conley2008a}.\\
  $^{i}$ The spectrum used is the classification spectrum reported by \citet{Ochner2012a}.\\
  $^{j}$ Amateur spectrum of 16 Feb. 2013 downloaded (at 03 Jun. 2014) from \url{http://users.northnet.com.au/~bohlsen/Nova/sn\_2013aa.htm}.\\  
\end{flushleft}
\end{table*}

\subsection{SN sample}
\label{sec:snsample}

\subsubsection{Sample selection}

As a basis for our study, we have considered SNe Ia for which the presence of blue-/redshifted narrow \NaI\,D components has been measured using intermediate- or high-resolution spectra by \citet{Sternberg2011a,Foley2012c} and \citet{Maguire2013a}. We selected 13 objects for our final modelling work (see Table \ref{tab:snsample}), demanding that the objects have an unambiguous classification of the \NaI\,D central wavelength shift, that a sufficient set of flux-calibrated low-resolution photospheric spectra is available (\ie usually a minimum of four spectra, with at least two of them before maximum light), and that they are considered `normal' SNe Ia \citep[\eg][]{Branch1993a,Filippenko1997a,Leibundgut2000a}.  Therefore, we have, \eg, excluded the SNe 1986G [strongly reddened, subluminous with \TiII\ lines -- \citet{Cristiani1992a}, \citet{Phillips1987a}, \citet{Ashall2016b}], 2006X [strongly reddened, high line velocities -- \citet{Wang2008a,Yamanaka2009a}], 2009ig [extremely high line velocities -- \citet{Foley2012a}] and LSQ12gdj [1991T-like -- \citet{Scalzo2014a}; \cf\ also \citet[][]{Phillips1992a,Jeffery1992a}]. We have also refrained from modelling PTF 11kx \citep{Dilday2012a}, for which time-variable \NaI\,D lines as well as other signs of CSM interaction have been found, but which shows a quite unusual spectrum almost without \SII\ $\lambda$$\lambda$5454,5640 after $B$ maximum. We note that, although we chose to concentrate on normal SNe Ia in this work, modelling peculiar objects will be an interesting subject for forthcoming studies.

\subsubsection{Division of our sample into `Na subsamples'}
\label{sec:snsample-subsamples}

We group our SNe Ia, following \citet{Sternberg2011a}, into four classes (`Na subsamples', see also Table \ref{tab:snsample}):
\begin{list}{$\bullet \ \ \ \ $}{\setlength{\labelwidth}{3.0cm} \setlength{\leftmargin}{0.65cm}}
 \item `blueshifted-Na' objects,
 \item `redshifted-Na' objects,
 \item one `single-Na' object,
 \item and `no-Na' objects.
\end{list}
In our terms, a `SN with blue-/redshifted Na' is an object with narrow \NaI\,D absorption components blue-/redshifted with respect to the host-galaxy \NaI\,D wavelength, where these SNe usually show their strongest narrow \NaI\,D absorption component. `No-Na' objects show no \NaI\,D line at all, while the `single-Na' object shows only one strong component at the host-galaxy \NaI\,D wavelength.

\citet{Maguire2013a} modified the classification scheme of \citet{Sternberg2011a}, by (i) measuring the \mbox{blue-/redshift} of the features with respect to an independently confirmed host-galaxy rest wavelength (obtained from sources other than the strongest \NaI\,D component), and (ii) statistically grouping the SNe into objects with blueshifted \NaI\,D components (`blueshifted-\NaI\,D' SNe -- regardless of redshifted components in their work), objects without blueshifted \NaI\,D components (`non-blueshifted-\NaI\,D' SNe) and objects without any \NaI\,D (`no-\NaI\,D' SNe). The SNe with blue- or redshifted Na we model in our work have shown \textit{no significant evidence} for a \NaI\,D component of the respective other sort. We have selected objects which, apart from rare cases (see remarks in Table \ref{tab:snsample}), are part of the respective class regardless of the classification method \citep{Sternberg2011a, Maguire2013a}.

\subsection{Data handling}
\label{sec:datahandling}

The photospheric spectra we use for the ejecta modelling have mostly been obtained within larger campaigns, such as the CFA SN follow-up \citep{Blondin2012a}, the PESSTO\footnote{\url{http://www.pessto.org}} survey \citep{Smartt2015a}, the Berkeley Supernova Ia program BSNIP \citep{Silverman2012a}, and the European SN Collaboration (EU RTN HPRN-CT-2002-00303). Table \ref{tab:snsample} lists the precise sources for the spectra object by object. Most of the data are publically available through the WISeREP repository\footnote{\url{http://www.weizmann.ac.il/astrophysics/wiserep/}} \citep{Yaron2012a}, and the PESSTO spectra are available through the ESO archive (see PESSTO web pages for details). The absolute calibration of the observed spectra has then been checked against photometry [from light-curve samples of \citet{Hicken2009a}, \citet{Ganeshalingam2010a}, \citet{Stritzinger2011a}, \citet{Brown2012a} and the papers cited in Table \ref{tab:snsample}] using \textsc{iraf}/\textsc{synphot}. 
In case deviations were found, the spectra affected were multiplied by a suitable polynomial function $f(\lambda)$ of degree $\leq$\,$2$ (with only a mild variation of $f(\lambda)$ with wavelength), respectively, such that the resulting sample of spectra is consistently flux-calibrated.

While we apply no reddening correction (correction for dust extinction) to the observed spectra, our spectral models are artificially reddened using a total extinction estimate (see Table \ref{tab:snsample}). The host-galaxy reddening is often subject to considerable uncertainty, especially in cases where we could only estimate the reddening from Na~ID line depths [\citet{Turatto2003a}; see \citet{Phillips2013a} and \citet{Poznanski2011a} for caveats]. Resulting uncertainties in our models are mitigated by the fact that we do not model highly reddened SNe~Ia. 

\subsection{Modelling method}
\label{sec:modellingmethod}

In order to analyse the photospheric spectra of our SN sample and derive ejecta properties, we use the abundance reconstruction method of \citet{Stehle2005a}. A modelling of nebular spectra and bolometric light curves, which is part of their `Abundance Tomography' approach, is not performed in this work, as for a number of our objects the respective data are not available. 

For modelling the photospheric spectra, we use a well-tested Monte-Carlo radiative transfer code (\citealt{Lucy1999a,Mazzali2000a,Stehle2005a}; for a short description see \eg \citealt{Mazzali2014a}) which calculates a synthetic spectrum from a given bolometric luminosity $L_\textrm{bol}$, time from explosion $t$, photosphere position/velocity\footnote{Owing to the homologous expansion of the ejecta setting in seconds to minutes after the explosion of a SN~Ia \citep[\eg][]{Ropke2005b}, $v$ and $r$ can be used interchangeably at any given point of time $t$ ($r=v\times t$).} $v_\textrm{ph}$, density profile $\rho(v)$, and abundance stratification $\mathbf{X}(v)$ (where $\mathbf{X}$ is the tuple of mass fractions of all relevant elements). 

To keep the parameter space for the models small enough to be thoroughly explored, $t$ and $\rho(v)$ follow a fixed parametrisation. The luminosity $L_\textrm{bol}$, some of the abundances $\mathbf{X}(v)$, and the photospheric velocity $v_\textrm{ph}$ are used as fitting parameters, which we optimise to match the observed spectra. In this fitting process, we infer a best-fit abundance stratification, from which we can assess differences between the observed SNe in terms of the chemical structure.

Before proceeding with the presentation of the models, we shortly discuss how we set the `fixed' code input parameters, and how we adjust the `free' input parameters when fitting an actual spectral sequence.

\subsubsection{Fixed parameters}

\begin{figure}   
   \centering
   \hspace*{-0.3cm}
   \includegraphics[angle=270,width=8.7cm]{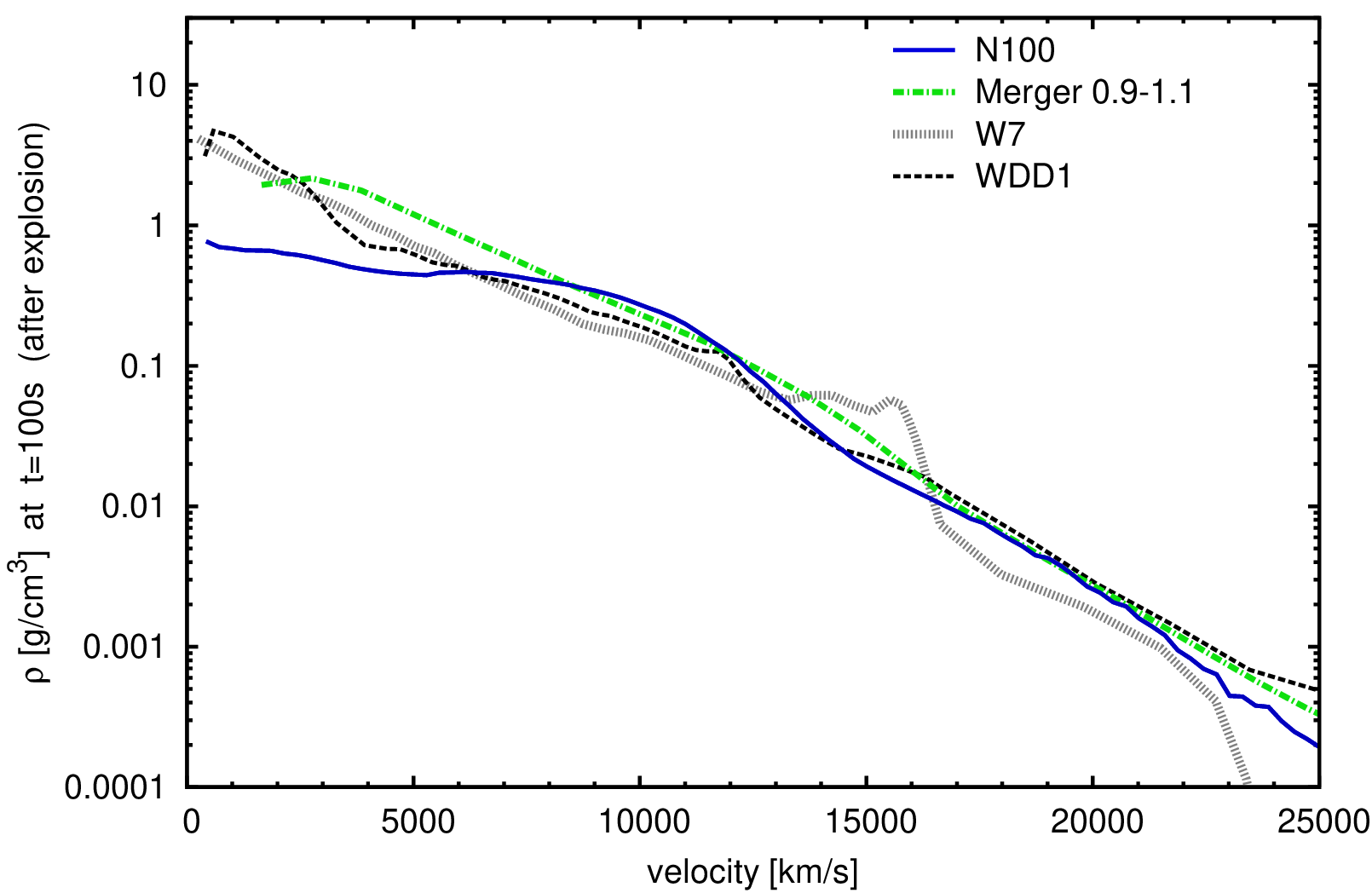}
   \caption{The N100 density profile \citep{Seitenzahl2013a} used as input data for our radiative transfer calculations [dark (blue), solid line]. For comparison, we show the density of a WD merger [0.9\Msun\ and 1.1\Msun\ C/O white dwarfs, \citet{Pakmor2012a} and \citet{Ropke2012a} -- light (green), dashed-dotted line]. We furthermore show models of \citet{Nomoto1984a,Iwamoto1999a} -- W7 (1D fast deflagration -- grey, broad/micro-dashed line) and WDD1 (1D delayed detonation -- black, dashed line), which have been used in many previous spectral modelling studies.}
   \label{fig:densitymodels}
\end{figure}

\begin{table*}
\begin{flushleft}
\caption{Prototypical abundance zones for our spectral models with their respective adjustable (\ie fitted within the given limits) and fixed abundances. The fixed abundances in zones I-VII (given to one significant digit for simplicity) are derived from the spatially-averaged nucleosynthesis of the N100 model \citep{Seitenzahl2013a} in different radial layers. Models other than N100 \citep[\eg][]{Iwamoto1999a,Gamezo2005a} have somewhat different chemical yields; however, qualitatively, the composition patterns remain similar. In each zone of our models, the abundances are chosen such that the spectra are optimally fit, \textit{under the constraint (see also text)} that the abundance pattern must match one of the zone patterns I-VII. Elements not mentioned are not unambiguously seen in the spectrum and are assumed not to be present (zero abundance). `Solar' abundances have been taken from \citet{Asplund2009a}.}
\vspace{-0.01cm}
\label{tab:nucleosynthesiszones}
\begin{tabular}{lccccccc}
\hline
$\!\!\!$ zone & $\!\!\!$ I $\!\!\!$ & $\!\!\!$ II $\!\!\!$ & $\!\!\!$ III $\!\!\!$ & $\!\!\!$ IV $\!\!\!$ & $\!\!\!$ V $\!\!\!$ & $\!\!\!$ VI $\!\!\!$ & $\!\!\!$ VII $\!\!\!$ \\ \hline
\multicolumn{8}{c}{\vspace{-0.2cm}} \\
$\!\!\!$ corresponding $v$ range $\!\!\!$ & $<$\,6000 & $\phantom{\textrm{1}}$6000\myto\quadbynine & 12000\myto\quadbynine & 13500\myto\quadbynine & 15000\myto\quadbynine   & 17000\myto\quadbynine  & $>$\,20000 $\!\!\!$ \\ 
$\!\!\!$  in N100 [\kms{}] &  & 12000$\phantom{\textrm{\myto}}$ & 13500$\phantom{\textrm{\myto}}$ & 15000$\phantom{\textrm{\myto}}$ & 17000$\phantom{\textrm{\myto}}$ & 20000$\phantom{\textrm{\myto}}$ & $\!\!\!$ \\ 
\multicolumn{8}{c}{\vspace{-0.2cm}} \\ \hline
\multicolumn{8}{c}{\vspace{-0.2cm}} \\
\multicolumn{8}{c}{fitted abundances; respective constraints for adjustment (if one abundance value is given: abundance fixed in respective zone)} \\
\multicolumn{8}{c}{\vspace{-0.2cm}} \\ \hline
\multicolumn{8}{c}{\vspace{-0.2cm}} \\
$\!\!\!$ $X(\textrm{C})$ [mass fraction, \%]    &  0.0$^a$                 & 0.0$^a$            &  0.0$^a$           & 0.0$^a$               & $\leq\textrm{0.3}$  & $\leq\textrm{3.0}$  & 12.0                        \\
$\!\!\!$ $X(\textrm{Si})$ [mass fraction, \%]   &  \multicolumn{2}{c}{$\!\!\!$adjusted $\quad\!\!\!$ such $\quad\!\!\!$ that $\quad\!\!\!$ $\sum_{i} X(i) = \textrm{100\%}$ $\!\!\!$}     & 30.0\myto40.0 & 40.0 & 40.0  & $\leq\textrm{30.0}$ & 5.0 or 1.0$^b$              \\
$\!\!\!$ $X(\textrm{Ca})^c$  [mass fraction, \%]   &  0.0$^a$              & $\leq\textrm{5.0}$ & $\leq\textrm{5.0}$ & $\leq\textrm{5.0}$    & $\leq\textrm{5.0}$  & $\leq\textrm{5.0}$        & $\leq\textrm{1.0}$ \\
$\!\!\!$ $X(\textrm{Fe}_0)^d$ [mass fr., \%]    &  3.0                & 10.0               &  7.0               & 1.0\myto5.0               & 0.1\myto1.0             & solar                     & solar                    \\
$\!\!\!$ $X(\textrm{\Nifs}_0)^d$ [mass fr., \%] &  70.0               & $\!\!\!$30.0\myto60.0$\!\!\!$  &  5.0\myto30.0     & $=X(\textrm{Fe}_0)$ & $=X(\textrm{Fe}_0)$/2 & $\leq X(\textrm{Fe}_0)$/10 & 0.0  \\
\multicolumn{8}{c}{\vspace{-0.2cm}} \\ \hline
\multicolumn{8}{c}{\vspace{-0.2cm}} \\
\multicolumn{8}{c}{fixed abundances} \\
\multicolumn{8}{c}{\vspace{-0.2cm}} \\ \hline
\multicolumn{8}{c}{\vspace{-0.2cm}} \\
$\!\!\!$ $X(\textrm{O})$ [mass fraction, \%]    & 0.0$^a$                       & 0.0$^a$           & \multicolumn{5}{c}{ $ - \quad$ adjusted $\qquad\!\!$ such $\qquad\!\!$ that $\qquad\!\!$ $\sum_{i} X(i) = \textrm{100\%}$ $\quad - $}                             \\
$\!\!\!$ $X(\textrm{Mg})$ [mass fraction, \%]   & 0.0$^a$                       & 0.0$^a$       &  0.0$^a$     & 1.0          & 4.0                & 9.0                 & $=X(\textrm{Si})$           \\
$\!\!\!$ $X(\textrm{S})$ [mass fraction, \%]    & $=X(\textrm{Si})/2$           & $=X(\textrm{Si})/2$ & $=X(\textrm{Si})/3$ & $=X(\textrm{Si})/3$ & $=X(\textrm{Si})/4$  & $=X(\textrm{Si})/4$ & $=X(\textrm{Si})/5$           \\
$\!\!\!$ $X(\textrm{Ti})$ [mass fraction, \%]   & 0.001                         & 0.002         &  0.003       & 0.007        & 0.005              & 0.001               & solar           \\
$\!\!\!$ $X(\textrm{V})$ [mass fraction, \%]    & 0.0300                        & 0.0400        &  0.0200      & 0.0020       & 0.0001             & solar               & solar           \\
$\!\!\!$ $X(\textrm{Cr})$ [mass fraction, \%]   & 0.70                          & 0.70          &  0.20        & 0.10         & 0.02               & solar               & solar           \\
\multicolumn{8}{c}{\vspace{-0.2cm}} \\ \hline
\end{tabular} 
\medskip \\
  $^{a}$ Set to 0.0 because abundance in this zone is too small to have an influence on synthetic spectra and integrated abundances.\\
  $^{b}$ 5.0\% when the early Si line of the SN extends to high velocities (blueshifts) indicating the presence of burning products at high velocity; 1.0\% otherwise. \\
  $^{c}$ Ca abundances are chosen larger than in the N100 model \citep[\cf][]{Seitenzahl2013a}, as otherwise the Ca lines in the spectra could not be well fitted. This problem may arise from uncertainties in the Ca atomic data for radiative transfer as well as in the exact nucleosynthesis; Ca abundances are not of major relevance for our further analysis in any case.\\
  $^{d}$ The abundances of Fe, Co and Ni in our models are assumed to be the sum of \Nifs\ and its decay chain products (\Cofs\ and \Fefs) on the one hand, and directly synthesised / progenitor Fe on the other hand (other contributions are can be neglected for spectral synthesis). Thus, they are conveniently given in terms of the \Nifs\ mass fraction at $t=0$ [$X($\Nifs$)_0$], the Fe abundance at $\,t=0$ [$X(\textrm{Fe})_0$], and the time from explosion onset $t$. \\
\end{flushleft}
\end{table*}

We use the spherically averaged density from the relatively recently published N100 three-dimensional delayed-detonation model (Figure \ref{fig:densitymodels}) of \citet{Seitenzahl2013a} as a fixed input density profile $\rho(v)$ reasonably representing SNe Ia. As \citet{Seitenzahl2013a} mention, out of their suite of explosion models, this model best matches the \Nifs\ production of typical SNe Ia [$\sim$\,0.6\Msun, \citet{Stritzinger2006b}]. Uncertainties in state-of-the-art explosion simulations (such as the unknown number and geometry of ignition points in Chandrasekhar-mass models), and the open question of the `typical' explosion scenario(s), clearly translate to some degree of uncertainty in our models. In order to give the reader a feeling for these uncertainties, we compare 
(Figure \ref{fig:densitymodels}) the N100 density profile to that of the `classical' W7 and WDD1 single-degenerate explosion models \citep{Nomoto1984a,Iwamoto1999a}, and to the density profile of a spherically averaged white dwarf merger [0.9\Msun\ and 1.1\Msun\ C/O white dwarfs, \citet{Pakmor2012a} and \citet{Ropke2012a}]. It is clear that the differences between the density profiles are significant, but not extreme in the region where photospheric spectra around maximum light form ($\sim$\,5000\myto25000\,\kms). The mass above 8000\,\kms, the velocity down to which we analyse the ejecta (see Sec. \ref{sec:models-fits} below), is almost equal in both models. Thus, we do not expect our results to depend much on the density profile used. We demonstrate this with an explicit test -- a repetition of our modelling with the merger density profile from \citet{Ropke2012a} -- in Appendix \ref{app:density-linefits-ddmodel}.

Having set the density profile in velocity space, the second input parameter to be fixed is the time from explosion onset to the epoch of a certain observed spectrum (where the latter is normally given with respect to $B$-band maximum by the observers). The time from explosion is calculated as
\begin{equation*}
 t = t_\textrm{obs.} + t_r
\end{equation*}
where $t_\textrm{obs.}$ is the observationally-determined time offset from $B$-band maximum, and $t_r$ is the $B$-band rise time which we calculate approximately as:
\begin{equation*}
 t_r = \textrm{18.9\,d} - \textrm{2.0\,d} \times \left( \frac{\Dm}{\textrm{mag}}-\textrm{1.17} \right).
\end{equation*}
We have chosen this formula as a linear approximation to the rise time in \Dm\ [the $B$-band light-curve decline during the first 15\,d directly after $B$ maximum, \citet{Phillips1993a}], assuming that SNe\,Ia with a faster decline also rise faster because of a smaller opacity. The parameters in the formula are tuned to obtain a rise time of 18.9\,d for SN~2002bo with $\Dm = \textrm{1.17}$ \citep[\ie to reproduce the values of][]{Benetti2004a} and to obtain a rise-time range of $\pm$\,1\,d for our SN sample (as given in Table \ref{tab:snsample}). This roughly corresponds to the spread found in light-curve samples of normal SNe Ia \citep[\eg][]{Conley2006a,Strovink2007a,Ganeshalingam2011a,Firth2015a}.

\subsubsection{Fitting parameters and fitting procedure}
\label{sec:fittingparameters}

Using the approach of \citet{Stehle2005a}, the radial abundance stratification is reconstructed from the outside inwards. We begin to optimise the abundances starting from typical `default' values similar to those inferred for SN~2002bo \citep{Stehle2005a}. For a more detailed description of the procedure, we refer to their original paper, but here we give at least a basic explanation.

To begin with, the earliest spectrum available for a respective SN is fit optimising the abundances in two separate zones \citep[\cf][]{Hachinger2013a}\footnote{In \citet{Hachinger2013a}, three separate zones were used as sufficient early-time data were available to avoid parameter degeneracies in the fitting process.}. One zone lies between the very outside of the simulated atmosphere (where negligible densities are reached, $v \sim \textrm{40000}\,\kms$) and 20000\,\kms\ -- in this zone, we assume a very weakly burned composition, with Fe-group elements only at the solar-metallicity level remaining from the progenitor (the exact metallicity level is irrelevant here, as we do not model the UV).
The second zone in which we optimise the abundances is between 20000\,\kms\ and the photospheric velocity $v_\textrm{ph}$ at the epoch of the first available spectrum. This photospheric velocity, let us call it `$v_\textrm{1}$', is determined such that the overall opacity of the atmosphere is reasonable, line velocities are matched and the overall colour of the observed spectrum can be reproduced. 
The luminosity  $L_\textrm{bol}$ is determined such that the overall flux level in the observed spectrum is matched by the synthetic spectrum. This quantity, together with the photospheric velocity $v_\textrm{ph}$ and the iron-group abundances largely determines the temperature structure within the model atmosphere. Thus, when any of these three parameters is changed, usually the abundance structure has to be somewhat re-adjusted so as to keep an optimum fit.

After matching the first spectrum, the fitting process is continued by finding abundances, $L_\textrm{bol}$ and $v_\textrm{ph}$ for the second spectrum. The new photospheric velocity, $v_\textrm{2}$, will be lower, corresponding to the recession of the photosphere into the intermediate ejecta with time. 
The abundances are then only fit between $v_\textrm{1}$ and $v_\textrm{2}$, while the values in the layers above $v_\textrm{1}$ are taken as given and remain fixed\footnote{An exception to this can be abundances which at an early epoch are not well determined because of degeneracies (\eg when two elements contribute to the same feature in the spectrum) or which only have moderate influences on the early-epoch fit. When the abundances chosen then prove incompatible with a later spectrum (effectively lifting the degeneracy), they are revised such as to fit all spectra in the time series equally well.}. 
This process is continued with later and later spectra (to determine the abundances between $v_\textrm{2}$ and $v_\textrm{3}$, and so on); we stop when we have modelled a spectrum 5\myto10 days after maximum. At this point, the photosphere has usually receded into zones rich in \Nifs, and the assumption that all radiation is generated below an assumed photosphere (which is implied by our modelling approach) becomes much less accurate.

As we model a large number of objects in the present study, we speed up the fitting process narrowing down the parameter space by only allowing abundance combinations `typical' for certain `burning zones' in the hydrodynamics/nucleosynthesis model N100 \citep{Seitenzahl2013a}. We note that 3D delayed-detonation explosion models like N100 do not exhibit the strictly-layered nucleosynthesis structure found in 1D simulations \citep{Thielemann1986a}. Instead, material burnt in the initial deflagration rises convectively; the strength of the deflagration and thus the `mixing' will depend on the somewhat uncertain ignition conditions. Finally, the deflagration turns into a detonation and practically the entire pre-expanded star undergoes nuclear reprocessing \citep{Seitenzahl2013a}. 
Despite all uncertainties, in Table \ref{tab:nucleosynthesiszones} we give a very approximate division of the ejecta of N100 into different zones with typical abundance patterns. Each abundance-fitting zone in our models (corresponding to a spectrum) is assigned the best-matching such zone, for which some abundances are considered fixed, and the others adjustable within limits (Table \ref{tab:nucleosynthesiszones}).
In the deepest layers of N100 ($v < \textrm{6000}$\,\kms), material burnt to nuclear statistical equilibrium (NSE) dominates -- mostly \Nifs\ [more neutron-rich material such as ${}^{54}$Fe is also produced, but partly ejected at higher velocities \citep{Seitenzahl2013a}]. Above this ejecta core, a zone with a very `mixed' composition is found; Fe-group elements are still dominant in the inner part of this zone, but intermediate-mass elements (IME such as Si and S, $\textrm{8} < Z \leq \textrm{20}$) dominate at its outer edge ($v \sim \textrm{12000}$\,\kms). Going still further outwards, zones dominated by IME follow, and above 17000\,\kms, finally O dominates as a burning product. The outermost envelope above \mbox{$\sim$\,20000\,\kms} is very weakly burned -- a feature directly implemented in our models.

\section{Models}
\label{sec:models}

Having laid out the modelling method, we discuss our model for SN~2007le (see Table \ref{tab:snsample}) as an example. This object is a representative `normal' SN~Ia (\Dm\,=\,1.0; no spectroscopic peculiarities), for which a series of (low-resolution) spectra with decent time coverage for the abundance analysis is available. High-resolution spectra of the object have revealed a blueshifted \NaI\,D component \citep{Sternberg2011a}. We first present the spectral fits and then discuss the abundance structure of the optimum-fit model.

Our model spectra and abundance structure plots for all other SNe from Table \ref{tab:snsample} can be downloaded as \textit{supplementary online material (part A)} to this preprint on the arXiv web site. Basic information on the models (luminosities, photospheric velocities, etc.) is included.

\subsection{Fits to the spectral sequence}
\label{sec:models-fits}

The spectral sequence for SN~2007le which we have modelled (Figure \ref{fig:spectra-2007le}) comprises six spectra from $-$10.4\,d to $+$9.6\,d (relative to $B$ maximum). This constitutes a fairly typical case with respect to epoch coverage, although there is a large gap (between $-$6.5\,d and $+$5.5\,d) around $B$ maximum. To reach a sufficient accuracy in the reconstruction of the abundance profile of SN~2007le, we fit three spectra at comparatively late epochs, which mitigates -- at least to a sufficient degree -- the uncertainties in abundance reconstruction caused by the gap.

Figure \ref{fig:spectra-2007le} shows the time sequence of models and observed spectra; in the top panel, we have given line identifications for a typical pre-maximum spectrum as we use it in our study. These line identifications are (to good approximation) valid for all our models; only some ten days after $B$ maximum, the main contributions to the features change, which reflects in changes in the spectral shape \citep[see also \eg][]{Stehle2005a}.

At the earliest two epochs, the model fits are very good, except for typical problems reproducing the extreme blue wings of the strongest lines (Ca H\&K, \SiII\ $\lambda$6355). These absorption wings, so-called high-velocity features (HVF; \eg \citealt{Mazzali2005a}, \citealt{Childress2014a}), are supposedly caused by an enhanced electron density in the outermost ejecta, the cause of which may be a density bump or mixing of the ejecta with H-rich material lost by the progenitor system \citep[\cf][]{Tanaka2008a}. In any case, this material has a small net mass (such that it does not contribute significantly to integrated abundances) and -- apart from the formation of the high-velocity components -- also a small influence on the spectrum. We thus merely ignore it in our study here.

The fits to the pre-maximum spectra allow for a robust abundance diagnosis of the IME-dominated outer ejecta (\cf\ line identifications in Figure \ref{fig:spectra-2007le}). The features dominated by Ca, Mg, S, Si and Fe lines are fit well overall. Some deficit of flux around $\sim$\,4000\,\AA\ may be attributed to a weak re-emission part of the \CaII\ H\&K P-Cygni profile in the model (corresponding to the missing high-velocity absorption). The mixed \FeII/\FeIII/\MgII\ absorption trough around 4300\,\AA\ is somewhat too shallow, probably owing to the low Mg mass fractions we use following the explosion models (Table \ref{tab:nucleosynthesiszones}). These problems are not expected to significantly affect our analysis in Section \ref{sec:analysis-discussion}, where we consider only major abundances (Fe-group, Si, O).

Our analysis is expected to be sufficiently reliable down to a velocity of  $\sim$\,8000\,\kms, where the abundances of iron-group elements increase in most SNe. Such a photospheric velocity is expected shortly after $B$ maximum. In mass space, the region analysed corresponds to 0.9\Msun\ of material (assuming the N100 density profile). The remaining core can only be well analysed using nebular spectra \citep[\eg][]{Kozma2005a,Mazzali2007a}. Correspondingly, after maximum light (Figure \ref{fig:spectra-2007le}, lowermost three panels) the fit quality degrades somewhat. Some lines are too deep in the models (especially \SII\ $\lambda$5640 -- here, part of the mismatch may be due to relatively large S abundances assumed following N100, see Table \ref{tab:nucleosynthesiszones}), and beyond $\sim$\,6500\,\AA, the model spectra show some flux excess. The latter is expected when assuming all energy generation to happen below a sharp photosphere and the photosphere to emit a black body \citep[\cf][]{Stehle2005a,Hachinger2013a}. Spectra later than a few days after $B$ maximum are therefore mainly useful as a rough consistency check, showing whether the best-fit composition found for the outermost layers is compatible with all data available.

\subsection{Abundances}

\begin{figure*}   
  \centering
  \includegraphics[angle=270,width=13.5cm]{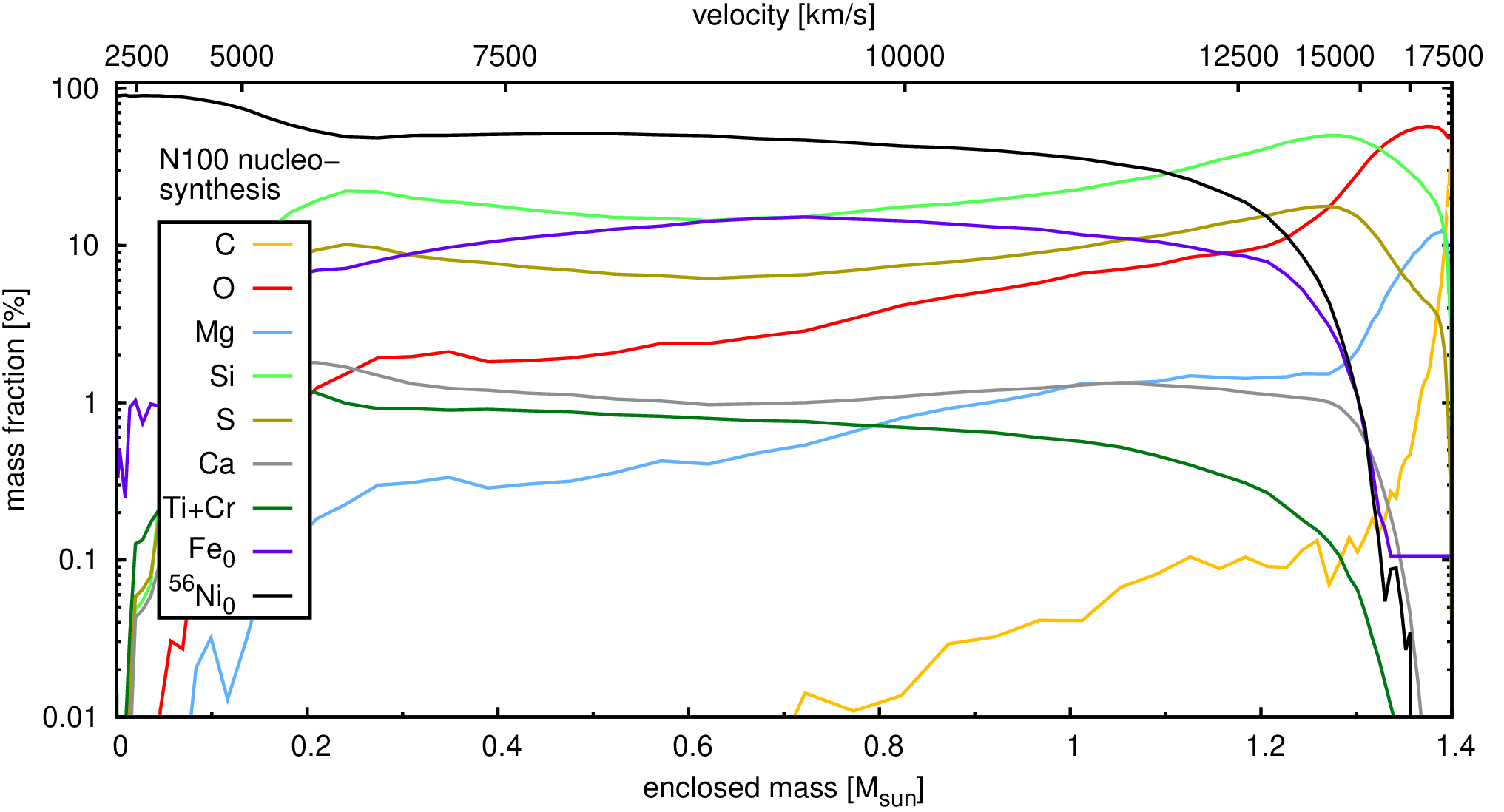}
  \\[0.15cm]
  \includegraphics[angle=270,width=13.5cm]{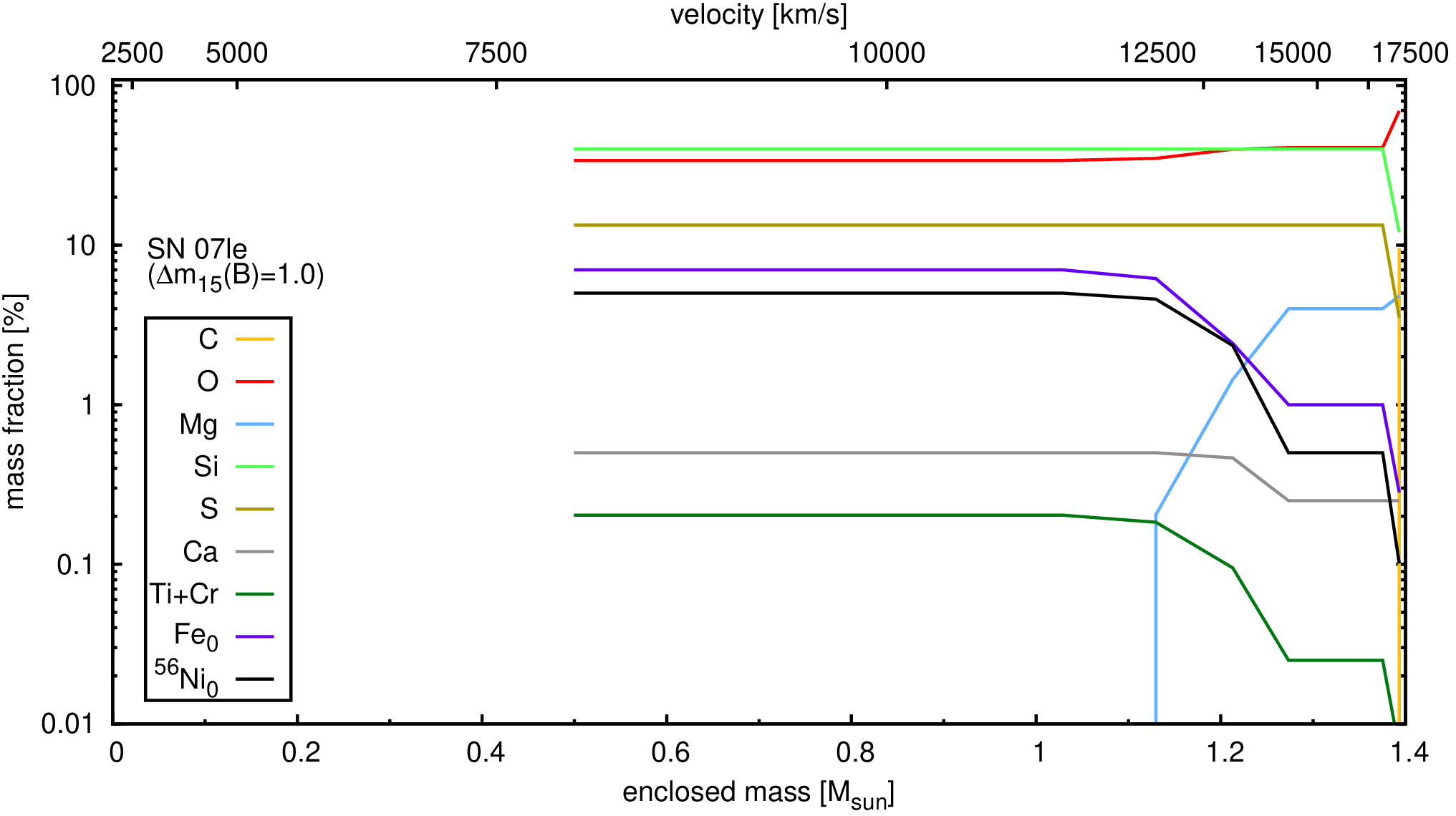}
  \\[-0.35cm]
  \includegraphics[angle=270,width=13.5cm]{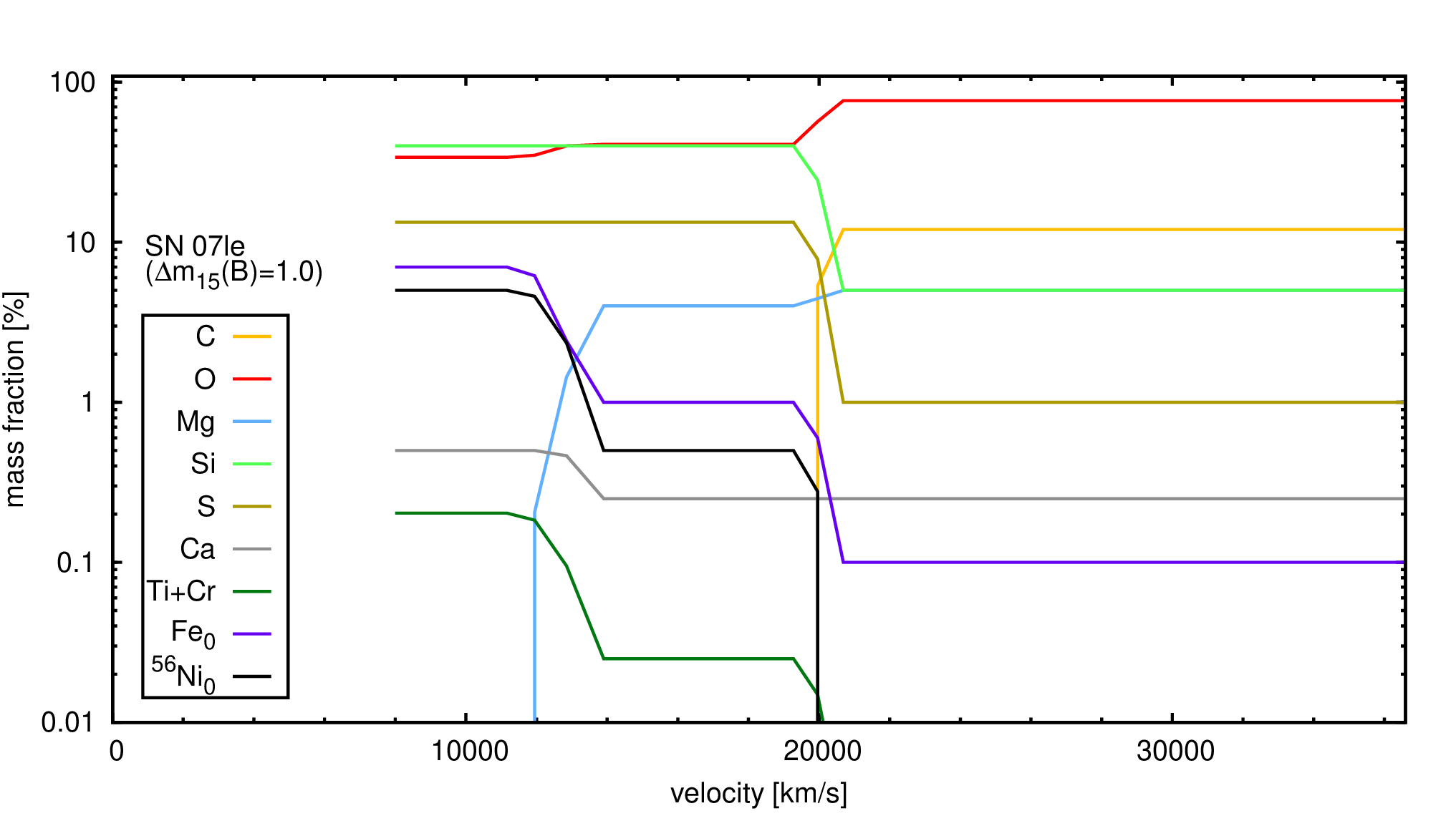}
  \caption{Abundance structure of N100 nucleosynthesis calculations \citep[][\textit{top panel}, plotted in mass space]{Seitenzahl2013a}, compared to our tomography of SN~2007le based on the N100 density profile (\textit{middle panel}: in mass space; \textit{lower panel}: in velocity space). The Ni\,/\,Co\,/\,Fe abundances are given in terms of the mass fractions of \Nifs\ and stable Fe at $t=\textrm{0}$ [$X(^{56}\textrm{Ni}_0)$, $X(\textrm{Fe}_0)$]; in our spectral models, no stable Ni or Co and no radioactive Fe are assumed to be present (as explosion models produce only amounts insignificant for photospheric spectra).}
  \label{fig:abundances-2007le}
\end{figure*}

The abundances we find for SN~2007le down to 8000\,\kms\ are shown in Figure \ref{fig:abundances-2007le}, together with the chemical structure obtained in the N100 simulation \citep{Seitenzahl2013a}, which can be considered a normally luminous `standard' SN~Ia model. 

\begin{figure*}   
   \centering
   \includegraphics[width=14.5cm]{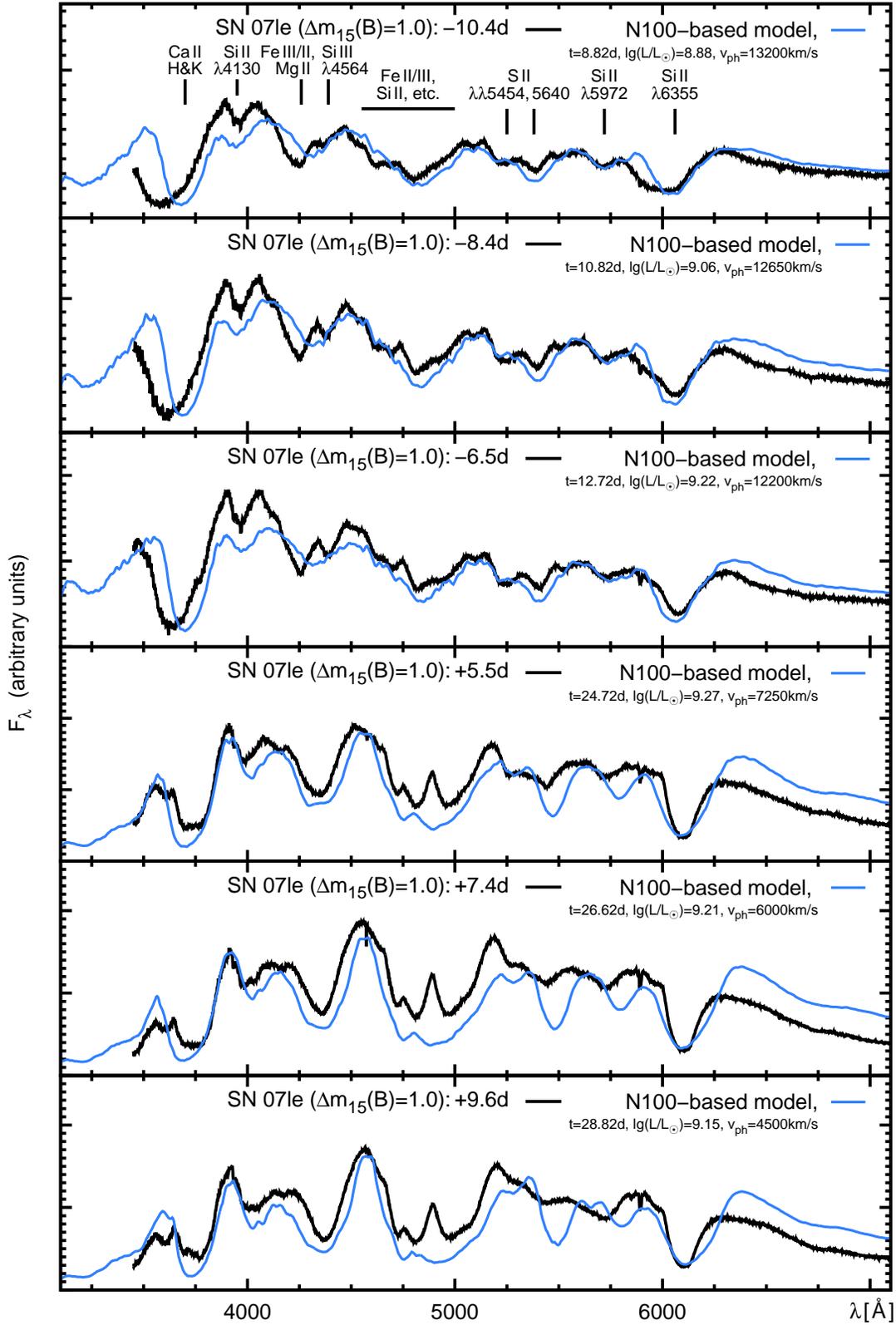}
   \caption{Model sequence for SN~2007le (blue lines). The observed (low-resolution) spectra (black lines) are plotted for comparison; line identifications are given for the most prominent features in the earliest spectrum. Observations and models are shown de-redshifted (\ie in the SN's rest frame); no correction for reddening has, however, been applied to the observed spectra [the models have thus been reddened by $E(B-V)_\textrm{total}=\textrm{0.22}$\,{}mag, applying a CCM extinction law with $R_V=\textrm{3.1}$ \citep{Cardelli1989a}].}
   \label{fig:spectra-2007le}
   \afterpage{\clearpage}
\end{figure*}

It is evident that SN~2007le does not show a \Nifs\ distribution extending as much out as in N100; this seems to be replaced by intermediate-mass elements. Therefore, the total mass of \Nifs\ in our model -- even if assuming a \Nifs-dominated core -- is $\lesssim$\,0.4\Msun, corresponding to a sub-luminous explosion with a bolometric luminosity of $\log\left(\frac{L}{\Lsun}\right)\lesssim 9.3$ \citep[equation 1 of][]{Stritzinger2006b}. This is plausible, given that our models some days before/after maximum light (Figure \ref{fig:spectra-2007le}) have a luminosity of $\log\left(\frac{L}{\Lsun}\right) \sim 9.2$ (with the spectral flux scale matching the data available), and that typical SNe\,Ia show a luminosity variation of about 0.1\,dex in this period \citep[\eg][]{Pastorello2007a}.

The reason why SN~2007le has shown a slow light-curve decline [\Dm\,$=$\,1.0, Table \ref{tab:snsample}] for such a low-luminosity object is not exactly clear. We speculate, however, that the relatively flat abundance gradients in the SN (implying that IME and NSE elements are found up to the very outer layers) may favour photon trapping inside the envelope and thus slow down the evolution of the light curve \citep[\cf][]{Mazzali2006a}.

\section{Analysis and discussion}
\label{sec:analysis-discussion}

In this section, we quantitatively compare models for SNe with blueshifted \NaI\,D lines to models for other SNe. First, we evaluate photospheric velocities in the models (Section \ref{sec:vphevolution}), which give hints about the overall opacity of the envelope (which, in turn, correlates with the composition). Afterwards, we show the integrated chemical yields of major species (Fe-group, Si, O) in the layers modelled (Sections \ref{sec:shallowfeg} and \ref{sec:oxygensilicon}). We conclude with a discussion of our findings in the context of the SN~Ia progenitor problem (Section \ref{sec:discussion}).

When evaluating model properties, we group our only single-Na object, SN~2002cr, with the redshifted-Na objects: both redshifted-Na and single-Na objects lack Na components from CSM (and show ISM lines at the same time), which means they should come from similar progenitor systems.

\subsection{Envelope opacity -- evolution of $v_\textrm{ph}$ with time}
\label{sec:vphevolution}

First, we search for evidence for a different opacity of the envelope, which may correspond to differences in the composition (or also the density structure), by looking at the time evolution of photospheric velocity $v_\textrm{ph}$ in our models (Figure \ref{fig:vphevolution}). For larger opacity, normally corresponding to a larger content of burned material [in particular of Fe-group elements, \cf \citet[][]{Sauer2006b}], the photosphere (in the sense of the layer where -- on average over all wavelengths -- an optical depth of one is reached) is expected to be further out in the envelope. Hints of a larger $v_\textrm{ph}$ among blueshifted-Na SNe may have indeed been found by measuring \SiII\ $\lambda$6355 Doppler velocities (blueshifts) as a proxy for $v_\textrm{ph}$ (\citealt{Foley2012c}, but see \citealt{Maguire2013a}).

The photospheric velocities in our models are highest among the blueshifted-Na SNe, and lowest among the no-Na SNe (with a noteworthy mean difference of 1000\myto{}2000\,\kms), indicating that blueshifted-Na SNe contain more opaque material in the outer layers. The redshifted-Na subsample comes to lie in between blueshifted-Na and no-Na objects, and thus shows no significant difference to the blueshifted-Na subsample or the no-Na subsample: we consider the photospheric velocities we find for each SN in this work to carry an error of $\pm$\,1000\,\kms, induced by uncertainties in the data (such as uncertain distances and reddening values) and the modelling approach\footnote{We estimate the uncertainty in the fitting process to be some hundreds of \kms\ ($\pm$). However, the majority of the SNe modelled here have not been studied in much detail. Thus, some data suffer from less well known distance and reddening, or from wavelength-dependent errors in flux calibration. This translates into an additional uncertainty to our final model parameters.}.

\citet{Benetti2005a} furthermore found that SNe Ia can observationally be grouped according to the \textit{time evolution} of their \SiII\ $\lambda$6355 Doppler velocities (which approximate the photospheric velocity). \citet{Tanaka2008a} modelled high-velocity-gradient [HVG, meaning a fast time evolution -- \citet{Benetti2005a}] and low-velocity-gradient (LVG) SNe and found that HVG objects have larger absolute photospheric velocities in addition to their large velocity gradient, corresponding probably to higher Fe-group mass fractions in the outer layers. The curves in Figure \ref{fig:vphevolution} show absolute velocity differences, but no apparent systematic difference in the gradients between the Na subsamples. In order to understand this in detail, however, one would need all our objects to have a better-sampled spectral time series (and thus model series) from early on to well past maximum [\cf \eg\ the data of \citet{Benetti2005a}].

\subsection{Mass of Fe-group material in the outer ejecta}
\label{sec:shallowfeg}

\begin{figure}   
   \centering
   \hspace*{-0.3cm}
   \includegraphics[angle=270,width=8.7cm]{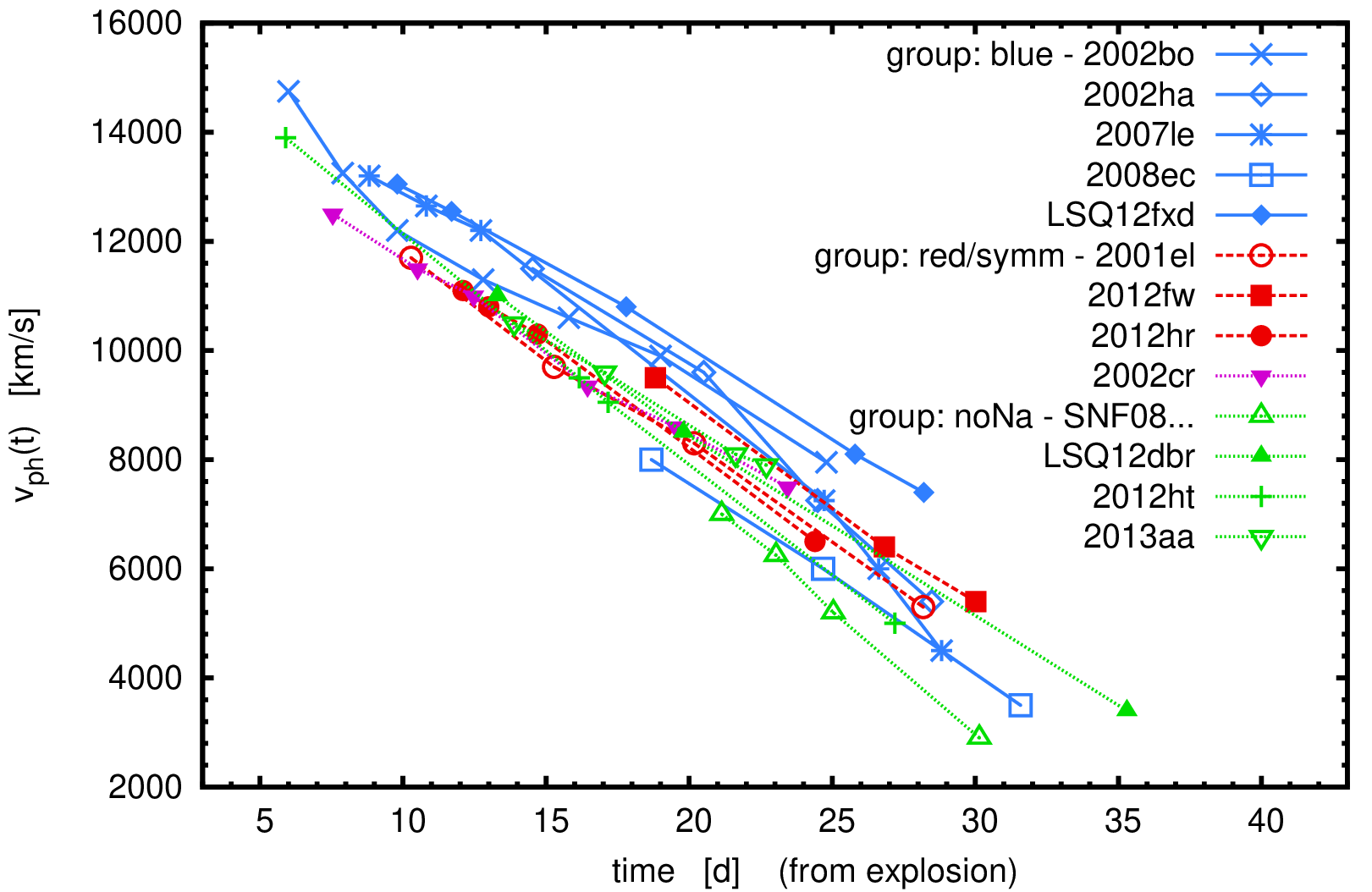}
   \caption{Evolution of photospheric velocities of our spectral models with the time passed since explosion onset. The curve style (solid/blue; dashed/red; dots/green) corresponds to the narrow Na line behaviour; different point types are used for every object.}
   \label{fig:vphevolution}
   \centering
   \hspace*{-0.3cm}
   \includegraphics[angle=270,width=8.7cm]{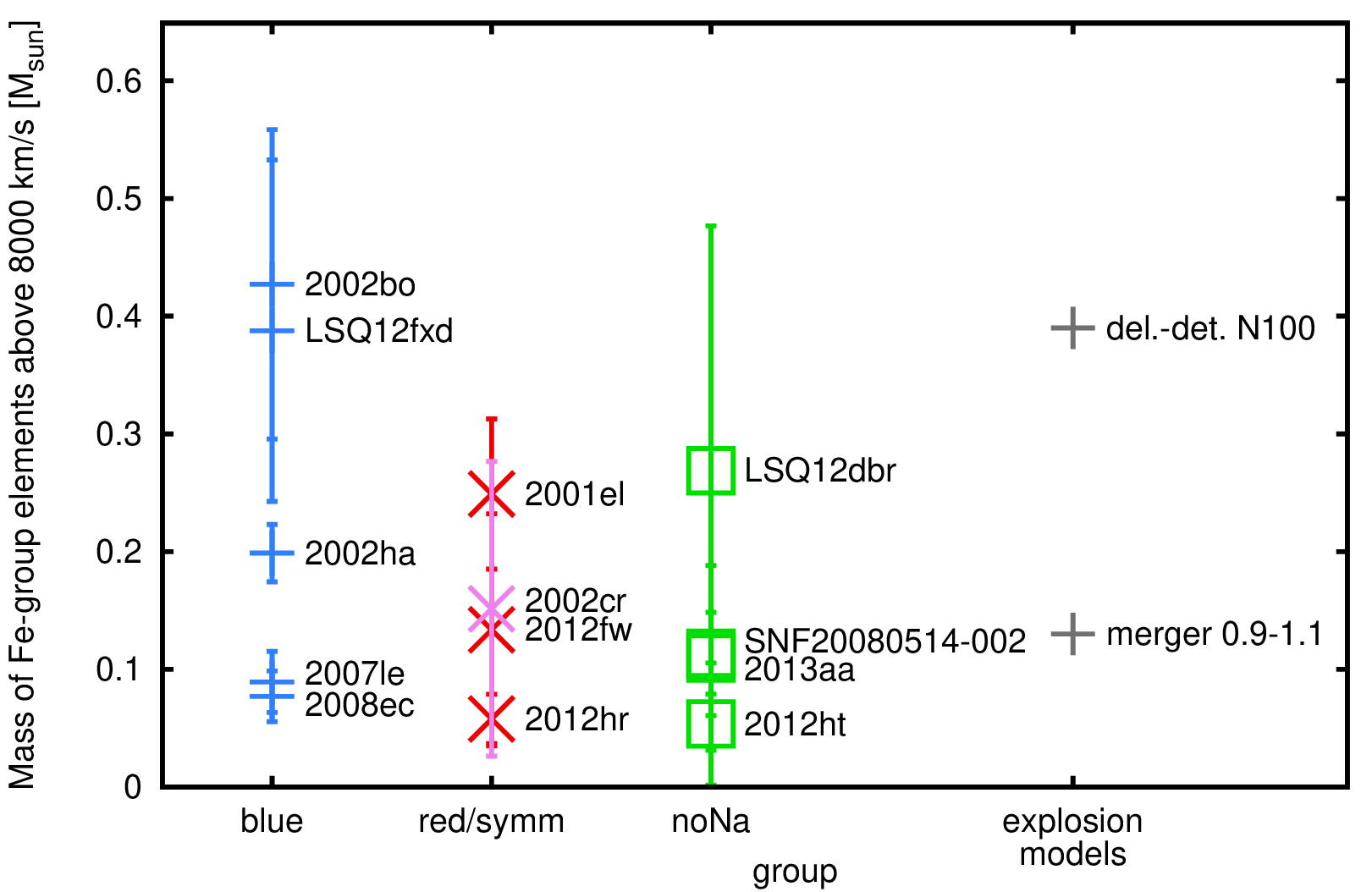}
   \caption{Mass of Fe-group elements above 8000\,\kms\ in our spectral models, which are grouped according to the narrow Na line behaviour of the corresponding SN (left three columns). The error bars (see also text) correspond to the error introduced by (mainly statistical) uncertainties in the photospheric velocity (and thus the abundance zone borders) which we estimate to be $\pm$\,1000\,\kms. For comparison, we plot the corresponding masses within the hydrodynamical explosion models of \citet{Ropke2012a} (right column).}
   \label{fig:nimasses}
\end{figure}

As a second model property to investigate, we consider the integrated Fe-group abundance. Our modelling approach allows us to determine the Fe-group content in the outer layers (above 8000\,\kms) from the strength of Fe lines (which have contributions from directly-synthesised Fe and from decayed \Nifs) and from the near-UV flux level of the SN (where all Fe-group elements, including the minor species, block the flux). We emphasise that there is some degeneracy between the effects of Fe, \Nifs, and Ti/V/Cr on the spectra as long as neither UV spectra nor very early spectra (where \Nifs/\Cofs\ have not decayed yet) and light curves are analysed \citep[\cf$\!$][]{Hachinger2013a}. As laid out in Section \ref{sec:modellingmethod}, we have avoided ambiguities in this respect by using Fe-group abundance mixtures typical for certain burning zones of a SN~Ia.

We plot (Figure \ref{fig:nimasses}, left three columns) the mass of Fe-group elements ($Z > \textrm{20}$) in our spectral models integrated from 8000\,\kms\ (down to which all our models have abundance information) outwards. The photospheric velocities in our models, giving the abundance zoning of the envelope (\cf\ Section \ref{sec:fittingparameters}) carry an uncertainty of $\pm$\,1000\,\kms\ (\cf Section \ref{sec:vphevolution}). This is a dominant source of error in the derived \textit{integrated} abundances which depend quite strongly on the zone borders. We have thus chosen to add error bars in Figure \ref{fig:nimasses} which are generated by determining the abundances with all abundance zone borders moved outwards/inwards (simultaneously) by 1000\,\kms. For comparison, Figure \ref{fig:nimasses} (right column) shows the Fe-group abundances within the ab-initio explosion/nucleosynthesis models of \citet{Ropke2012a} on the right-hand side.

The Fe-group masses in the outer ejecta of blueshifted-Na SNe are a bit higher on average than those of other SNe. The mean values (with equal weight for each object) for the three Na subsamples and their error estimates (obtained by quadratic propagation) are:
\begin{equation*}
M(\textrm{Fe-group})_\textrm{blue} = \textrm{0.24}\pm\textrm{0.04}\Msun,
\end{equation*}
\begin{equation*}
M(\textrm{Fe-group})_\textrm{red/symm} = \textrm{0.15}\pm\textrm{0.04}\Msun\ \ \ \textrm{and}
\end{equation*}
\begin{equation*}
M(\textrm{Fe-group})_\textrm{no-Na} = \textrm{0.15}\pm\textrm{0.06}\Msun.
\end{equation*}

As one recognises easily, the variation between the subsamples is practically within the error bars and thus hardly significant in a statistical sense. However, we remark that the differences we see between blueshifted-Na SNe and the other subsamples (both with lower Fe-group abundances), are in line with the larger opacity we suspected blueshifted-Na SNe to have after the $v_\textrm{ph}$ analysis (Section \ref{sec:vphevolution}).

Compared to the \Nifs\ mass differences among explosion models, the variations we find are somewhat smaller. We note that the low value of Fe-group elements in the analysed layers of no-Na SNe may simply reflect a lower-than-average \Nifs\ production in these SNe (rather than only a difference in the distribution, as expected between equally-luminous double-degenerate and single-degenerate explosions). The lower value for the no-Na subsample would then be in line with the typically lower luminosities in the explosion sites of these objects, which are mainly early-type galaxies \citep{Hamuy2000a,Gallagher2008a,Maguire2013a,Ashall2016a,Henne2017a}. For redshifted-Na SNe, a more likely explanation would be `real' differences in the explosion properties (as luminosities are relatively similar among our redshifted-Na and blueshifted-Na objects). More objects should be investigated in the future in order to obtain statistically significant evidence.

\subsection{Mass of oxygen and silicon}
\label{sec:oxygensilicon}

\begin{figure}   
   \centering
   \hspace*{-0.3cm}\includegraphics[angle=270,width=8.7cm]{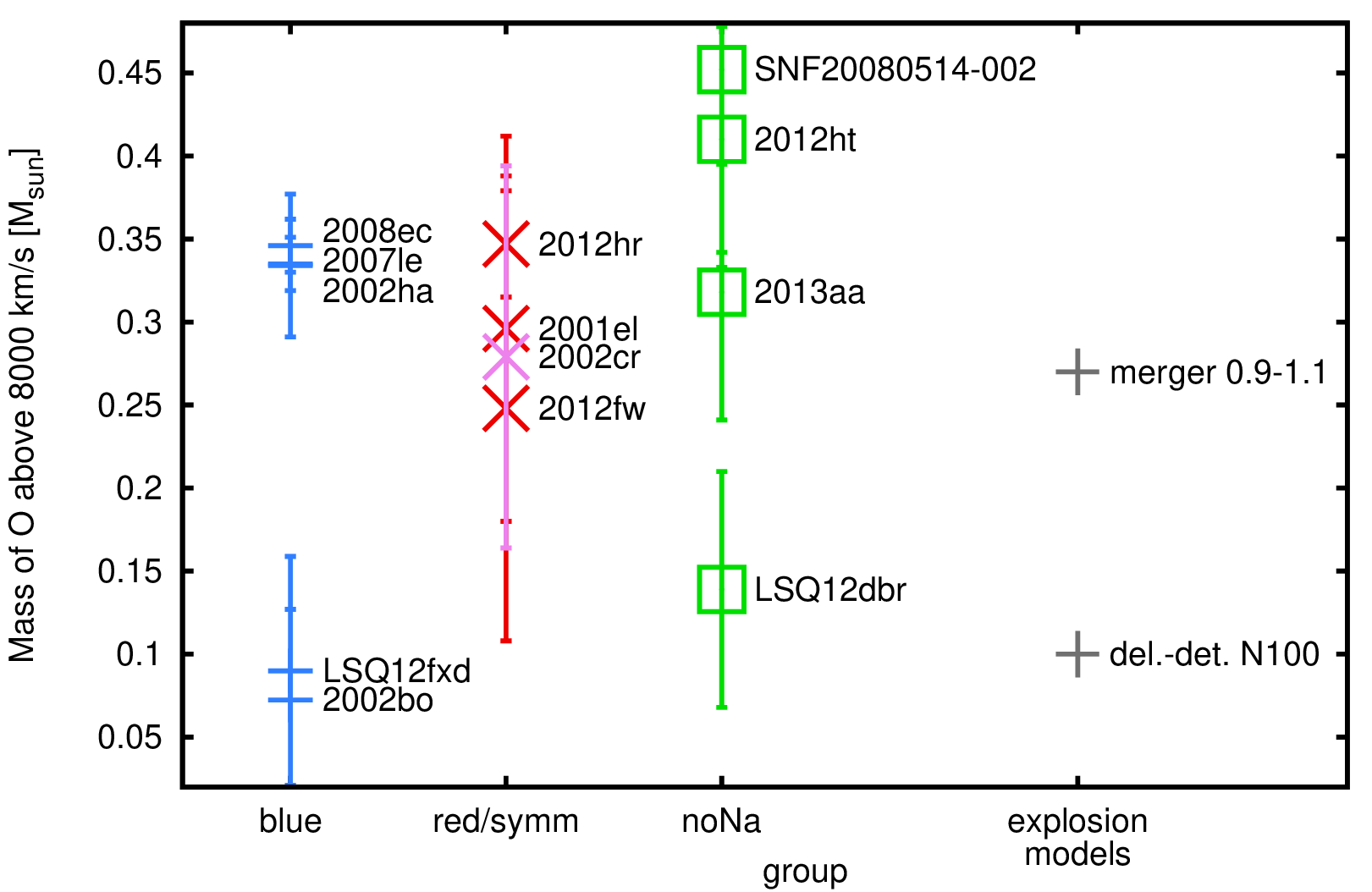}
   \hspace*{-0.3cm}\includegraphics[angle=270,width=8.7cm]{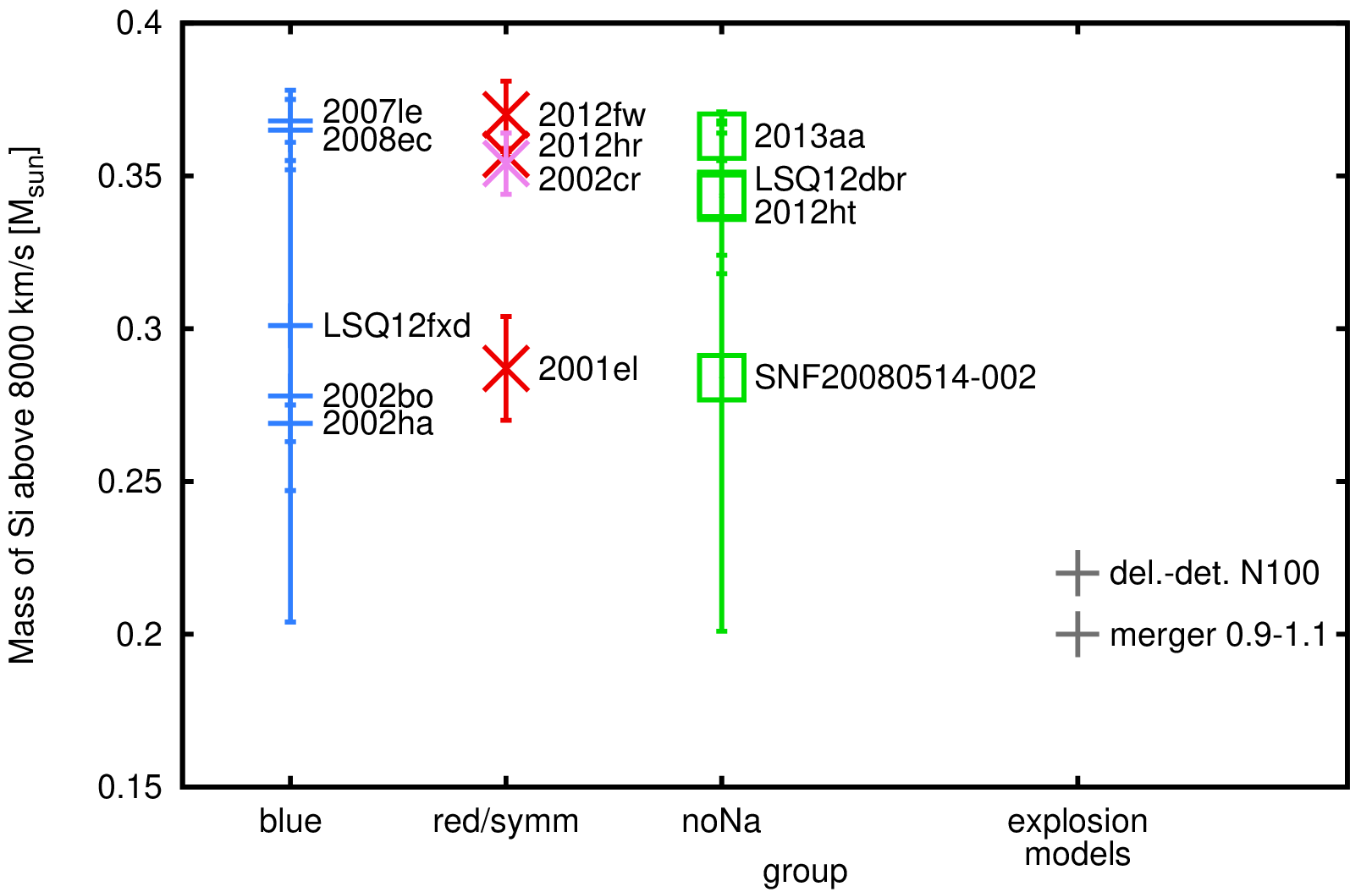}
   \caption{Mass of O (\textit{upper panel}) and Si (\textit{lower panel}) above 8000\,\kms\ in our models, which are grouped according to the narrow \NaI\,D line behaviour of the corresponding SN (left three columns, respectively). The error bars correspond to the error in $v_\textrm{ph}$ ($\pm$\,1000\,\kms, as in Figure \ref{fig:nimasses}). For comparison, we plot the corresponding masses within the hydrodynamical explosion/nucleosynthesis models of \citet{Ropke2012a} (right column within each panel). The variations in Si mass are actually relatively small (fine mass scale in lower panel) both among our SNe and the hydro models. The merger model `compensates' for its smaller Fe-group mass in the outer layers (Figure \ref{fig:nimasses}) by having a larger O abundance and a significant amount of unburned C ($\sim$0.1\Msun, not plotted). }
   \label{fig:o-si-mass}
\end{figure}

As a final quantitative criterion, we examine the O and Si content of the outer/intermediate layers of our models, down to 8000\,\kms. In Figure \ref{fig:o-si-mass}, we have plotted the yields of our different models, and of the explosion/nucleosynthesis models of \citet{Ropke2012a}.

We see that no-Na and redshifted-Na objects make up their deficit in shallow Fe-group material (Section \ref{sec:shallowfeg}) by having somewhat more O (Figure \ref{fig:o-si-mass}, upper panel) and a little bit more Si \citep[Figure \ref{fig:o-si-mass}, lower panel; see also][]{Mazzali2007a}. The difference in O mass between violent-merger and delayed-detonation explosion models is roughly in line with the observed differences (higher mean O mass in redshifted-Na and no-Na SNe). In Si mass, the explosion models only show an insignificant difference to one another (with absolute values somewhat lower than the observations).

Owing to the relatively large scatter, the differences between the groups with respect to O\footnote{It does not come as a big surprise that a particularly large scatter is present among our O abundances: apart from the uncertainties discussed in the sections above, the O mass of an optimised spectral model is a less-precisely determined quantity owing to weak effect of oxygen on the spectrum -- \OI\ $\lambda$7773, the only O feature, is often contaminated by other elements' lines (mainly Mg, Si).} and Si are less clear than those we are able to see in the Fe-group statistics or in the photospheric velocities. Thus, we will mainly focus on interpreting the latter differences in the following.

\subsection{Implications on SN Ia progenitors}
\label{sec:discussion}

We have found small differences between SNe displaying blueshifted-Na and other SNe; these differences exceed the typical error margins we have calculated by little or remain within them. In photospheric velocities, there is a difference of about 1000-2000\,\kms\ between blueshifted-Na SNe and no-Na SNe (at any epoch), with redshifted-Na SNe lying in between. The lower photospheric velocities in no-Na (and redshifted-Na) SNe point towards a lower envelope opacity and probably a less efficient nucleosynthesis. Accordingly, blueshifted-Na SNe show a Fe-group abundance higher by about 0.1\Msun\ in the outer layers with respect to both the other subsamples.

We will now try and interpret these findings in the context of SN~Ia progenitors.

\subsubsection{Theoretical expectations: differences from double- to single-degenerate explosions}
\label{sec:possibledifferences}

Single- and double-degenerate explosions are expected to show significant differences not only in the spatial density distribution [with possible asymmetries in double degenerates, \citet{Pakmor2010a}], but also in the abundance structure \citep[\cf][]{Ropke2012a}. The (violent) white dwarf merger models of \citet{Pakmor2012a} and \citet{Ropke2012a}, as a promising example of double-degenerate models, produce relatively large quantities of O and Si (which largely results from the burning of the secondary white dwarf) in the core of the ejecta, but also in the outer layers. Fe-group material is produced burning the dense core of the more massive white dwarf, and thus will be found in the centre of the ejecta. 3D delayed-detonation explosions, as a realistic single-degenerate model for SNe Ia \citep{Seitenzahl2013a}, have less O and Si in the centre. Fe-group material is found in their core, but also further outwards (\cf\ Figure \ref{fig:abundances-2007le}) because of the turbulent combustion in the initial deflagration phase. Quantifying the different Fe-group content in the models above 8000\,\kms\ (\cf\ Figure \ref{fig:nimasses}), one obtains a value of $\sim$\,0.4\Msun\ for N100 \citep{Seitenzahl2013a} and a value of $\sim$\,0.1\Msun\ for the 0.9\myto{}1.1\Msun\ merger of \citet{Pakmor2012a} and \citet{Ropke2012a}.

\subsubsection{Observational expectations: are different explosion types strongly linked to different Na subsamples?}
\label{sec:probabilityexpectations}

When analysing our Na subsamples for possible systematic differences caused by different progenitors, we have to ask the basic question how clearly SNe of one or another Na subsample would represent SNe from one or another explosion mechanism.

For a simplified discussion, we assume two kinds of progenitor systems to our observational subsamples -- one of which is able to produce CSM (`CSM producers')\footnote{Usually, `CSM producers' are believed to be single degenerates, but see \citet{Phillips2013a}, \citet{Shen2013a} and \citet{Raskin2013a}. We therefore apply a neutral naming here.} and the other not (`CSM-free systems'). CSM-producer objects need not always show blueshifted narrow Na lines, as the CSM may be asymmetric or very dilute. CSM-free systems, of course, can not show CSM Na lines. All systems may show random \NaI\,D components from the ISM. 

It then turns out that the probabilities of finding CSM producers in the different Na subsamples depend crucially on what fraction of SNe~Ia actually come from CSM producers. In the \textit{supplementary online material (part B)}, we have calculated these probabilities for two illustrative example scenarios (30\% CSM producers or 80\% CSM producers within all SNe~Ia), assuming the observed Na-subsample sizes of \citep[\eg][]{Maguire2013a} to be representative. We have obtained the following results:
\begin{list}{$\bullet \ \ \ \ $}{\setlength{\labelwidth}{3.0cm} \setlength{\leftmargin}{0.65cm}}
 \item `Scenario 1': For a CSM-producer fraction of 30\%, the majority (60\%) of blueshifted-Na SNe will be CSM producers, while the no-Na and redshifted-Na subsamples will be practically devoid of these systems (contribution $\ll \textrm{10}\%$). This means, the Na subsamples can be mapped relatively clearly to different progenitor types.
 \item `Scenario 2': If 80\% of all SNe~Ia are from CSM producers, practically all blueshifted-Na SNe (90\%) are CSM producers, but the no-Na and redshifted-Na subsamples are also essentially composed of these objects (70\%), \ie\ the difference between the Na subsamples becomes insignificant. In other words, if almost all SNe come from CSM-producer systems, all Na subsamples will have equal SN properties, and the fact that many SNe do not show CSM lines will simply be due to intrinsic effects within the explosion scenario (asymmetries etc., as mentioned).
\end{list}

\subsubsection{How many / which progenitor channels are at work?}
\label{sec:progenitorchannels}

From our study, which found slight differences between the Na subsamples, we can not definitely answer the question whether `normal' SNe Ia originate from one or two fundamental types of progenitor systems. If further studies, in the line of \citet{Sternberg2011a}, \citet{Foley2012c}, \citet{Maguire2013a} and this work, however continue to find little difference especially among redshifted-Na and blueshifted-Na SNe, we may suggest `Scenario 2' from the previous subsection to hold: most SNe Ia are due to the same explosion mechanism(s) that is/are able to produce CSM lines. In this context, it is important to clarify whether not only single-degenerate systems \citep{Booth2016a}, but also double-degenerate ones \citep{Phillips2013a,Shen2013a,Raskin2013a} can generate CSM signatures. It is also important to understand why SNe Ia in early-type galaxies tend to be fainter than average and display no narrow \NaI\,D lines. If all SNe Ia (with redshifted/blueshifted, but also without Na lines) come from the same progenitor systems, the correlation between early-type host and lack of \NaI\,D might be explained by what \citet{Booth2016a} suggest: \NaI\,D lines may only be visible if the CSM interacts with a relatively dense ISM.

\section{Conclusions}
\label{sec:conclusions}

Using radiative-transfer modelling, we have performed an abundance analysis of a sample of SNe\,Ia for which information on narrow \NaI\,D absorption lines (in high-resolution spectra), as well as a decent spectral time sequence (low resolution, but good time coverage and sampling), is available. Our aim was to identify possible differences between SNe\,Ia with and without narrow blueshifted \NaI\,D absorption features, and how these may relate to their explosion mechanisms. Narrow \NaI\,D absorption lines blueshifted with respect to the SN rest frame are -- at least in some cases -- supposedly formed within progenitor material along the line of sight towards the observer. 

The most important differences we find between our `Na subsamples' are in photospheric velocity (where blueshifted-Na SNe show the highest values, followed by redshifted-Na SNe and no-Na SNe), and in the Fe-group content of the outer layers (where blueshifted-Na SNe show higher values than both redshifted-Na and no-Na SNe). However, all these differences are practically within error bars, such that with our sample and methods firm conclusions on SN Ia progenitors can not yet be derived.

We note that our results are compatible with those of \citet{Maguire2013a}, who find little difference among the subsamples showing \NaI\,D lines (redshifted vs. blueshifted), but some larger difference between blueshifted-Na SNe and no-Na objects. Comparing the latter two subsamples, we see different photospheric velocities and Fe abundances within the outer ejecta. These correspond to a smaller overall Fe-group content in no-Na SNe, reflecting the fact that these no-Na SNe almost exclusively occur in early-type galaxies, where SNe Ia tend to have narrower light curves and a smaller luminosities \citep[\eg][]{Hamuy2000a}. Between blueshifted- and redshifted-Na SNe, we find a marginally significant difference in the Fe-group content of the outer layers. This may be compatible with systematic differences in the abundance stratification, possibly compatible with different explosion mechanisms. However, as long as we have only small-number statistics, this remains speculation. Mild variations, as we see them, may rather well be in line with a `unified' SN scenario for blue- and redshifted-Na objects.

In order to clarify whether one or multiple SN~Ia scenarios are at work, the differences between the Na subsamples should be further examined using much larger samples of the order of 100 objects. This becomes particularly clear looking at the expectation values of finding different explosion types within the subsamples. Unless CSM-producing systems are a minority of all SNe Ia ($\sim \textrm{30}\%$), they will contribute a significant number of objects to all Na subsamples, making no-Na, redshifted-Na and blueshifted-Na SNe hardly distinguishable by any data-analysis method applied to small samples.

As more data become available, also theoretical SN Ia models should evolve, allowing for more precise checks whether observational results are compatible with them. In particular, it has to be theoretically clarified whether only single-degenerate binaries, or also double-degenerate systems can generate a CSM producing narrow \NaI\,D components in the SN spectrum. Furthermore, a possible correlation of the CSM production in these systems with other progenitor/SN properties (such as metallicity) should be studied. Meanwhile, an analysis of peculiar SNe Ia, which we have excluded from our study and which in quite some cases have shown blueshifted \NaI\,D lines, may help to shed further light on the progenitors of SNe Ia.

\section*{ACKNOWLEDGEMENTS}
S.H., F.K.R. and A.G.-Y. have been supported by a Minerva ARCHES award of the German Ministry of Education and Research (BMBF). F.K.R. has been additionally supported by the Emmy Noether Programme (RO 3676/1-1), and A.G.-Y. acknowledges support by the ISF and the Lord Sieff of Brimpton Fund. K.M. acknowledges support from the STFC through an Ernest Rutherford Fellowship. M.S. acknowledges support from the Royal Society and EU/FP7 ERC grant n$^{\rm o}$ 615929, and S.T. acknowledges support by the TRR 33 `The Dark Universe' of the German Research Foundation. S.J.S. acknowledges funding from the EU/FP7 (2007-2013) ERC Grant n$^{\rm o}$ 291222.

We have used data from the NASA\,/\,IPAC Extragalactic Database (NED, \href{http://nedwww.ipac.caltech.edu}{http://nedwww.ipac.caltech.edu}, operated by the Jet Propulsion Laboratory, California Institute of Technology, under contract with the National Aeronautics and Space Administration). Furthermore, we have made use of the Weizmann Interactive Supernova data REPository [WISeREP -- \href{http://wiserep.weizmann.ac.il}{http://wiserep.weizmann.ac.il}, \citet{Yaron2012a}] and of the Lyon-Meudon Extragalactic Database (LEDA -- \href{http://http://leda.univ-lyon1.fr}{http://leda.univ-lyon1.fr}, supplied by the LEDA team at the Centre de Recherche Astronomique de Lyon, Observatoire de Lyon).

Larger parts of this research are based on observations collected at the European Organisation for Astronomical Research in the Southern Hemisphere, Chile as part of PESSTO (the Public ESO Spectroscopic Survey for Transient Objects) ESO program ID 188.D-3003, and on the CfA Supernova Archive, which is funded in part by the National Science Foundation through grant AST 0907903. We also made use of a classification spectrum from the Asiago Supernova classification programme of the INAF-OAPD SN group.

\textsc{IRAF} with \textsc{STSDAS / SYNPHOT} and \textsc{TABLES} version 3.6 has been used for spectrophotometry and data handling in this paper. \textsc{IRAF} - Image Reduction and Analysis Facility - is an astronomical data reduction software. It is distributed by the National Optical Astronomy Observatory (NOAO, \href{http://iraf.noao.edu}{http://iraf.noao.edu}, operated by AURA, Inc., under contract with the National Science Foundation). \textsc{STSDAS / SYNPHOT} and \textsc{TABLES} are \textsc{IRAF} packages provided by the Space Telescope Science Institute (STSCI, \href{http://www.stsci.edu}{http://www.stsci.edu}, operated by AURA for NASA).

\bibliographystyle{2013-mn2e}
\bibliography{astrohaetschi.bib}

\clearpage
\appendix

\section{The assumption of the N100 density in our models}
\label{app:density-linefits-ddmodel}

As laid out in Section \ref{sec:modellingmethod}, we assume the density profile of the N100 model (\citealt{Seitenzahl2013a}) for our abundance analyses. This model represents the explosion of a Chandrasekhar-mass C/O WDs as a delayed detonation. Thus, it likely corresponds to the single-degenerate path of evolution to a SN Ia. Here, we test whether this introduces  systematic errors into our results for the chemical yields. We show results based on a modified version of our models, which is based on the spherically averaged density of the violent-merger explosion model of \citet{Pakmor2012a} and \citet{Ropke2012a}. The latter model represents a SN Ia resulting from a double-degenerate system of two C/O white dwarfs with 0.9\Msun\ and 1.1\Msun\ (see also plot in our main text -- Section \ref{sec:method-data}, Figure \ref{fig:densitymodels}).

\begin{figure}   
   \centering
   \vspace{-0.6cm}
   \hspace*{-0.3cm}\includegraphics[angle=270,width=8.0cm]{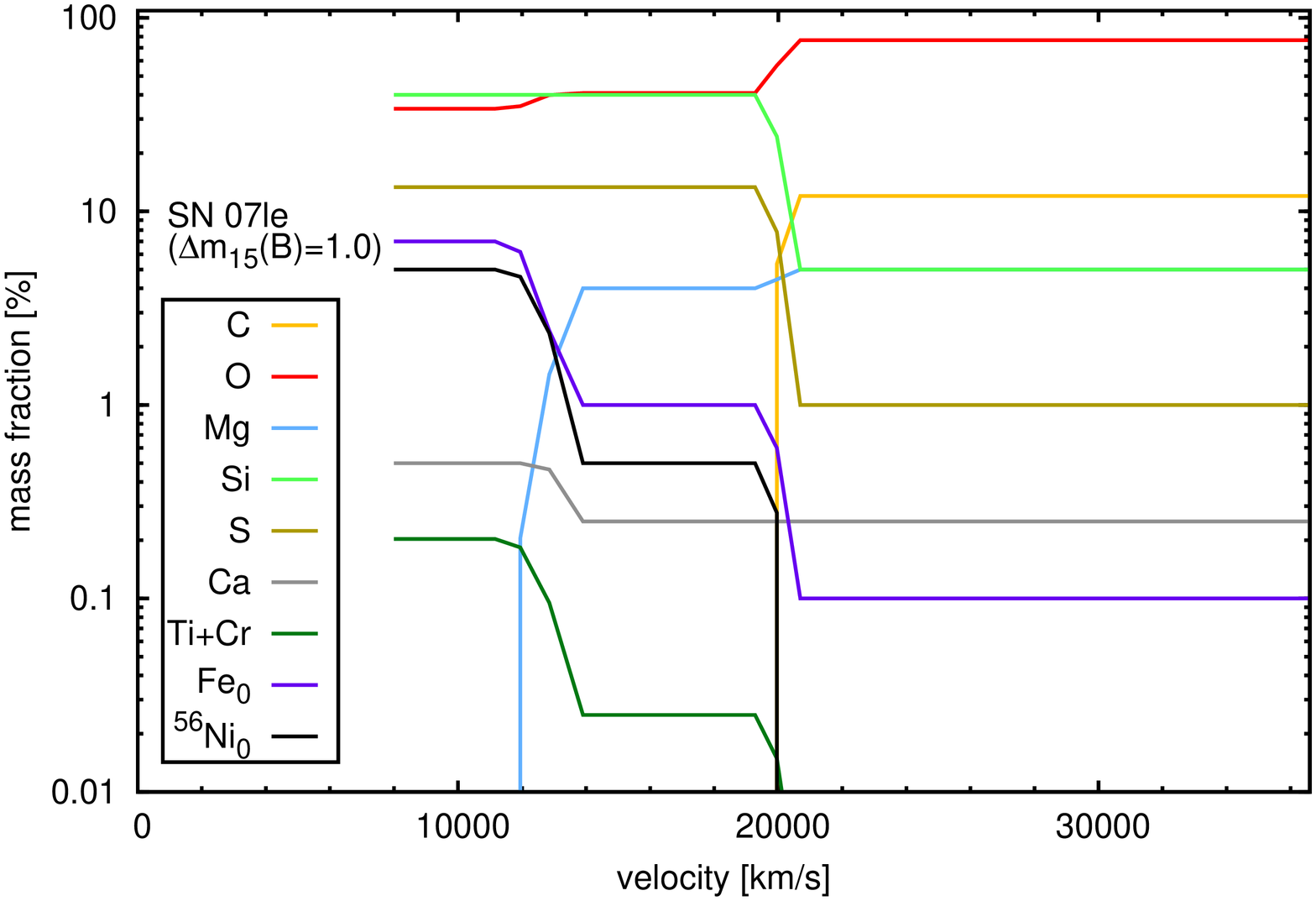}\\[-0.5cm]
   \hspace*{-0.3cm}\includegraphics[angle=270,width=8.0cm]{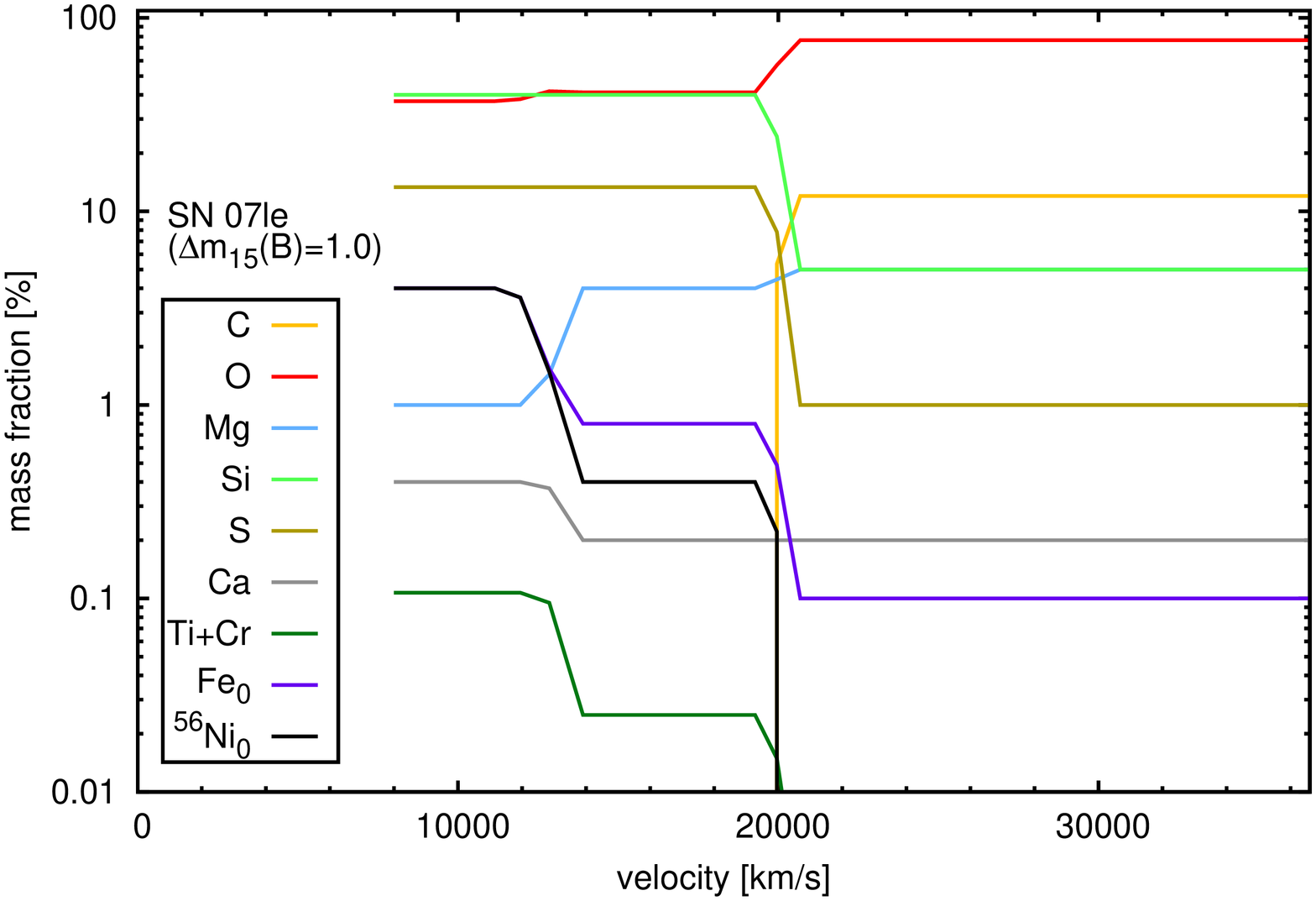}
   \caption{\textit{Upper panel:} Abundance distribution of our (N100-based) standard model for 2007le (already shown in Figure \ref{fig:abundances-2007le}). \textit{Lower panel:} Abundance distribution of the test model based on the merger density.}
   \label{fig:density-abund-merger-compare}
   \centering
   \vspace{0.1cm}
   \hspace*{0.0cm}\includegraphics[angle=270,width=7.5cm]{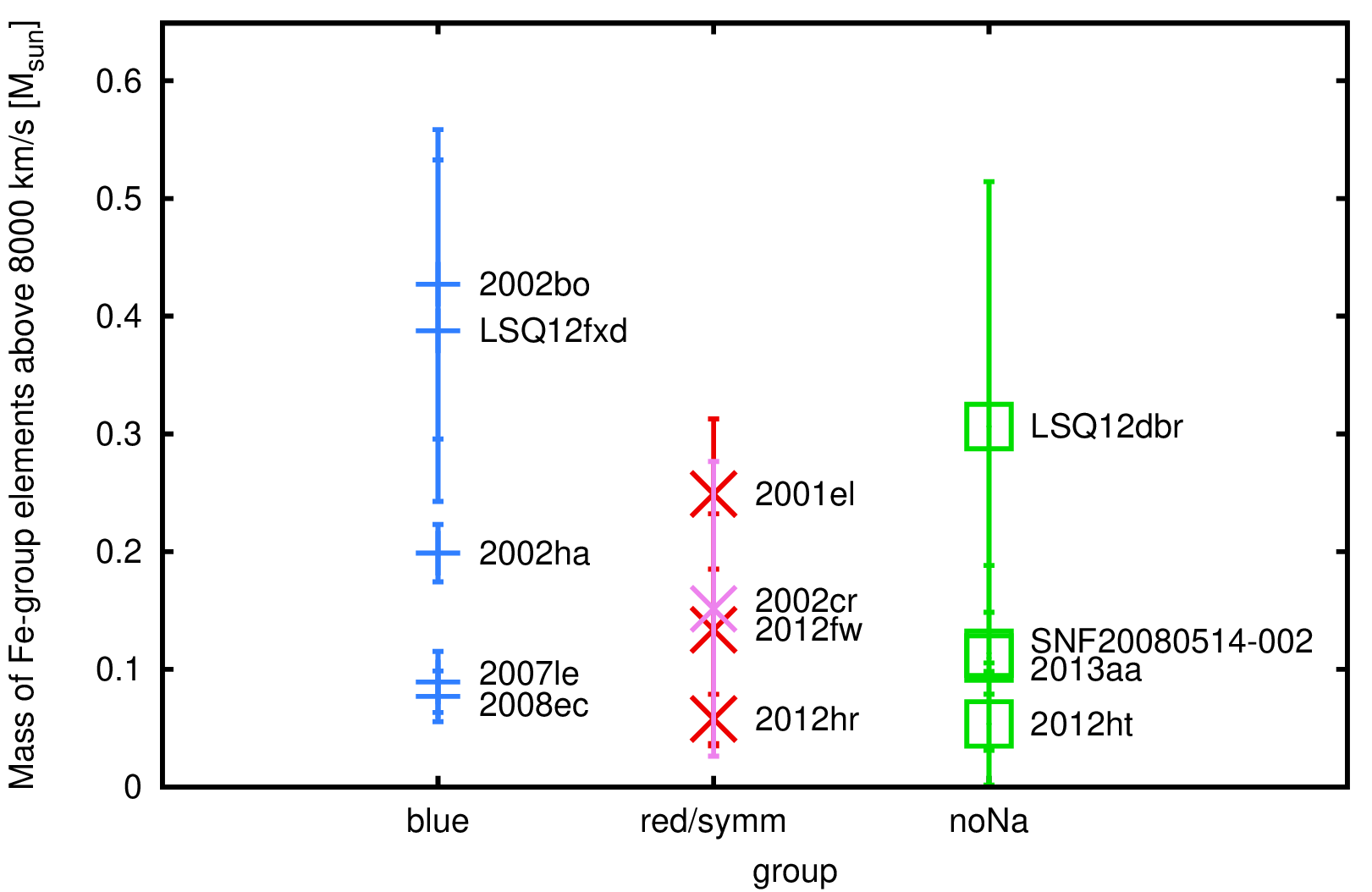}\\[-0.0cm]
   \hspace*{0.0cm}\includegraphics[angle=270,width=7.5cm]{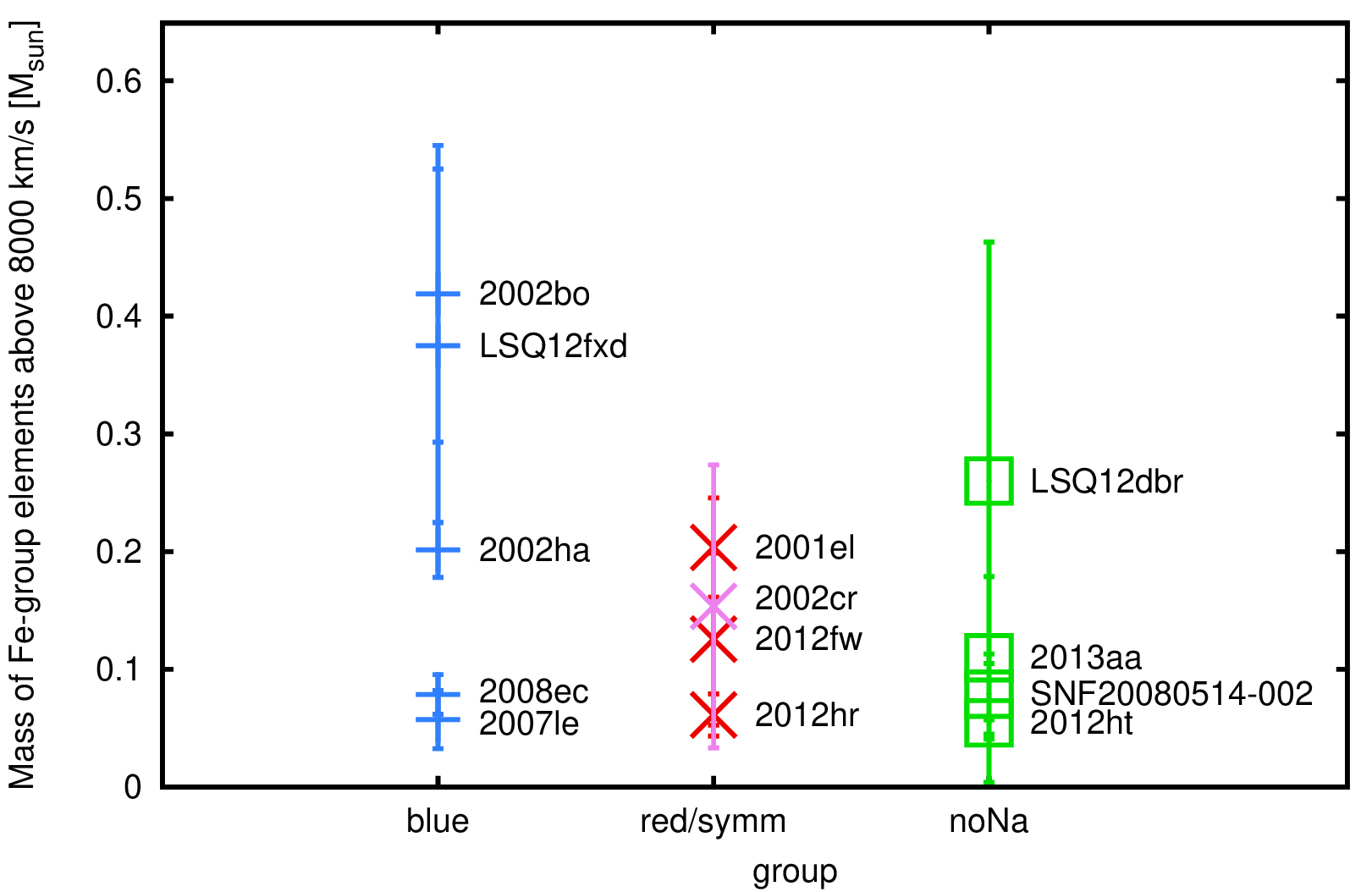}
   \caption{Mass of Fe-group elements above 8000\,\kms\ in our models based on N100 (\textit{upper panel}, which is a copy of Figure \ref{fig:nimasses} of the main text), and in our test models based on the 0.9\myto1.1\Msun\ merger density (\textit{lower panel}). The differences in density between N100 and the merger model have no pronounced effect on the chemical yields of our models.}
   \label{fig:nimasses-merger-compare}
\end{figure}

In the upper and lower panel of Figure \ref{fig:density-abund-merger-compare}, we give the N100- and the merger-based abundance reconstruction for SN~2007le as an example case. We had to make only minor modifications to abundances/photospheric velocities to obtain an optimum fit with the merger model. The mass in both the merger and the delayed-detonation model above $v=\textrm{8000}$\,\kms\ (\ie in the envelope zone we have modelled) is $\sim$\,0.9\Msun, such that also the integrated chemical yields are similar.

Therefore, when repeating our analyses of Section \ref{sec:analysis-discussion} with the merger-based models for all our SNe, we obtain only insignificant differences. An example for this is shown in Figure \ref{fig:nimasses-merger-compare}. Here, the mass of Fe-group elements above $v=\textrm{8000}$\,\kms\ we derive is shown for all SNe, first for our `standard models' (upper panel, see also Section \ref{sec:shallowfeg}) and then for our merger-based models (lower panel). Realising that the changes are minor, we are confident that our conclusions would not be changed if we would use the merger model as a basis for all our analysis or for part of it. 

We caution that both our N100- and our merger-based models are calculated with 1-D-averaged density profiles of the respective explosion models. In terms of fit quality, the results we obtain with these averaged density profiles are sometimes not as good as results obtained with a density from 1-D explosion models such as W7 or WDD1 \citep{Nomoto1984a,Iwamoto1999a}. The discrepancies between models and observations, which may be due to the hydrodynamical models themselves or to the 1-D averaging, correspond to an uncertainty in our abundance results which we estimate to be within the error bars shown in Section \ref{sec:analysis-discussion}. In forthcoming studies with larger SN Ia samples, where the entire analysis reaches a finer (\ie\ statistically more significant) level, this source of error should however be thoroughly investigated.

\clearpage

\end{document}